\documentclass[twocolumn,times,floatfix]{aastex631} 
\usepackage{graphicx}
\usepackage{mathtools} 
\usepackage{natbib}
\usepackage{amsmath}
\usepackage{verbatim}
\usepackage{appendix}
\usepackage{multirow,tabularx}
\usepackage{outlines}
\usepackage[version=4]{mhchem} 
\usepackage{graphicx}
\usepackage{comment}
\usepackage{float}
\usepackage{colortbl}
\usepackage{xkcdcolors}

\usepackage[version=4]{mhchem} 

\newcommand{\um}{$\mathrm{\mu}$m}
\newcommand{\Ang}{\mbox{\normalfont\AA}}
\newcommand{\PIE}{Pre-Industrial Earth}
\newcommand{\ARE}{Archean Earth}
\newcommand{\LyA}{Lyman-$\alpha$}
\newcommand{\OsD}{$\mathrm{O(^1D)}$}
\newcommand{\Seff}{S$\mathrm{_{eff}}$}

\defcitealias{NASEM:2021_Astro2020Decadal}{NASEM 2021}

\received{\today}
\revised{\today}
\accepted{\today}
\submitjournal{ApJ}

\shortauthors{Davis et al.}

\begin{document}

\title{The Effects of M Star Age Dependent Ultraviolet Emission on Detecting and Interpreting Exoplanet Biosignatures}

\correspondingauthor{C. Evan Davis}
\email{c.evan.davis@ucsc.edu}

\author[0009-0005-3386-6091]{C. Evan Davis}
\affiliation{Department of Astronomy and Astrophysics, University of California Santa Cruz, 1156 High Street, Santa Cruz, California 95064, USA}

\author[0000-0002-1386-1710]{Victoria S. Meadows}
\affiliation{Astrobiology Group, SETI Institute, 339 Bernardo Ave, Suite 200, Mountain View, CA, 94043}
\affiliation{Department of Astronomy and Astrobiology Program, University of Washington, Box 351580, Seattle, Washington 98195, USA}
\affiliation{NASA NExSS Virtual Planetary Laboratory, Box 351580, University of Washington, Seattle, Washington 98195, USA}

\author[0000-0002-7260-5821]{Evgenya L. Shkolnik}
\affiliation{NASA NExSS Virtual Planetary Laboratory, Box 351580, University of Washington, Seattle, Washington 98195, USA}
\affiliation{School of Earth and Space Exploration, Arizona State University, Tempe, AZ 85281, USA}

\author[0000-0003-0429-9487]{Andrew P. Lincowski}
\affiliation{Department of Astronomy and Astrobiology Program, University of Washington, Box 351580, Seattle, Washington 98195, USA}
\affiliation{NASA NExSS Virtual Planetary Laboratory, Box 351580, University of Washington, Seattle, Washington 98195, USA}

 \author[0000-0002-1046-025X]{Sarah Peacock} 
\affiliation{University of Maryland, Baltimore County, MD 21250, USA} 
\affiliation{NASA Goddard Space Flight Center, Greenbelt, MD 20771, USA}
\email{sarah.r.peacock@nasa.gov}

\author[0000-0001-5646-6668]{R. O. Parke Loyd}
\affiliation{School of Earth and Space Exploration, Arizona State University, Tempe, AZ 85281, USA}

\author[0000-0002-6294-5937]{Adam C. Schneider}
\affil{United States Naval Observatory, Flagstaff Station, 10391 West Naval Observatory Rd., Flagstaff, AZ 86005, USA}

\author[0000-0002-7129-3002]{Travis Barman}
\affiliation{Department of Planetary Sciences, University of Arizona, Tucson, AZ 85721, USA}






\begin{abstract}

Given their abundance and observational advantages, M stars will arguably be the best candidates for characterizing and searching for biosignatures on terrestrial exoplanets in the near future.  However, photochemistry that can suppress or enhance key biosignature molecules in planetary atmospheres is primarily driven by UV flux from the host M star, which is influenced by stellar activity that decreases with age.  Here, we simulate \PIE-like and \ARE-like atmospheres around M4 and M8 stars from 650 Myr to 5 Gyr old.  We find that our \PIE{} atmospheres around 5 Gyr M stars have up to ten times more \ce{CH4} than those around 650 Myr M stars, producing 68 \% stronger methane bands in NIR transit spectroscopy.  Additionally, photochemical shielding from \ce{O2} in our \PIE{} atmospheres reduces the impact UV-driven photochemistry on composition, while the \ARE{} exhibits larger compositional changes due to weaker shielding from \ce{CO2}.  Lastly, enhanced \ce{CO2} photolysis, driven by the strong net UV flux and high Far/Near-UV ratios of 650 Myr and 1 Gyr M stars, cause our \ARE-like planets to produce up to 5.4 dex more \ce{O3} than when around 5 Gyr M stars.  The excess \ce{O3} causes the \ARE{} to become half as reflective in the 0.2-0.3 \um{} Hartley band feature in ultraviolet reflectance spectroscopy, which the Habitable Worlds Observatory may be sensitive to.  Without the context of the star's real-time, age-dependent UV radiation, this \ce{O3} feature could be misinterpreted as a proxy for low, biogenic \ce{O2}.

\end{abstract}



\section{Introduction} \label{sec:intro}



In the near term, M stars \citep[$0.075 \lesssim M \lesssim 0.6M_{\odot}$,][]{Benedict:2016_MdwarfMLR} will arguably be the best host candidates for characterizing resident terrestrial exoplanets and the search for life \citep[see reviews by][]{Shields:2016, Meadows:2018_PCb}.  M stars make up about 70\% of the local stellar neighbourhood \citep{Bochanski:2010}, and the conditions within their proto-planetary disks favor the formation of low mass planets with small semi-major axes \citep{Wu:2013_KeplerDynamics, Mills:2016_Kep223ResChain}.  As such, many have been found to host terrestrial exoplanets \citep{Anglada-Escude:2016_PCbDiscovery, Gillon:2017_T1Discovery, Luger:2017_T1hConfirm} in their habitable zones (HZ).  M stars are also less luminous than stars like the Sun, and planets in the HZs of M stars have orbital periods on the order of days \citep{Dressing:2015}.  The small radii of M stars also make their planets more amenable to transit studies, as the transit depth, $(R_p/R_*)^2$, is higher for these smaller stars.  These characteristics make the atmospheres of planets orbiting M stars accessible to both JWST \citep{Lustig-Yaeger:2019_T1JWST,Meadows:2023_JWSTBiosigs,Greene:2023_T1bJWST, Zieba:2023_T1cJWST} and ground-based ELTs \citep{Lopez-Morales:2019_GroundBasedO2,Currie:2023_GroundBasedO2}, and as such are prime targets for the characterization of resident terrestrial exoplanets in the HZ.


Observations of terrestrial exoplanets orbiting M stars with JWST have ushered in the era of terrestrial exoplanet atmosphere characterization and the search for biosignatures.  Recent studies on TRAPPIST-1 b with JWST are most consistent with the planet having little or no atmosphere \citep{Greene:2023_T1bJWST, Gillon:2026_T1bcNoThickAtm}.  For TRAPPIST-1 c, initial observations were consistent with an airless world, or modeled steam atmospheres of up to 3 bar, which were 1.7--1.8$\sigma$ from the data \citep{Zieba:2023_T1cJWST, Lincowski2023_T1cAtmComp}.  JWST thermal phase curve observations are also consistent with an airless world, and suggest that massive steam atmospheres are unlikely \citep{Gillon:2026_T1bcNoThickAtm}. However, tenuous \ce{O2}-dominated atmospheres with traces of greenhouse gases such as \ce{CO2} and \ce{H2O} are still consistent with the current data \citep{Gillon:2026_T1bcNoThickAtm}.  While \citet{Piaulet-Ghorayeb:2025_T1d} find that NIRSpec/PRISM spectra of TRAPPIST-1 d are consistent with stellar contamination and an airless transiting body, they are also marginally consistent (within 2$\sigma$) with thin Mars-like and cloudy Venus-like atmospheres.  Several studies have discussed the feasibility of TRAPPIST-1's habitable zone planets --- TRAPPIST-1 e and f --- possessing atmospheres, with some studies predicting secondary atmosphere retention depending on initial volatile inventory and sequestration \citep{Krissansen-Totton:2023_T1AtmRetention, Piaulet-Ghorayeb:2025_T1d, Gialluca:2026_AtmSweeps} while others predict complete atmospheric loss within a Gyr over a wide range of atmospheric compositions \citep{VanLooveren:2024_T1AtmLoss}, especially if heavier volatiles (like C, S, O, and N) are subject to direct hydrodynamic escape.  Indeed, recent JWST transit-transmission spectroscopy of TRAPPIST-1 e showed that spectral retrievals were most consistent with a flat line, which could be indicative of an airless body, a high mean-molecular-weight atmosphere, or a thick, high-altitude cloud deck \citep{Espinoza:2025_T1eJWST, Glidden:2025_T1eJWST}.

Characterization of M star terrestrial exoplanet atmospheres, should they exist, may be possible via transit transmission spectroscopy and secondary eclipse and phase curve measurements \citep[e.g.][]{Morley:2017, Krissansen-Totton:2018_T1eJWST, Lustig-Yaeger:2019_T1JWST, Kreidberg:2019,Greene:2023_T1bJWST}.  Ground-based Extremely Large Telescopes (ELTs) may also be capable of searching for \ce{O2} and other biosignature molecules, such as the \ce{CO2}/\ce{CH4} disequilibrium pair, in M star terrestrial exoplanet atmospheres, using both transmission and reflected light spectroscopy \citep{Snellen:2015, Lopez-Morales:2019_GroundBasedO2, Currie:2023_GroundBasedO2, Currie2025_DirectImagingDetectability}.  Further in the future, a space-based direct-imaging observatory, such as the Habitable Worlds Observatory prioritized by the Astro2020 Decadal Survey \citepalias{NASEM:2021_Astro2020Decadal}, may be sensitive to molecules like \ce{O2} and \ce{O3} \citep{Feng:2018_ReflectanceRetrieval} and will search for signs of these molecules on dozens of planets orbiting F, G, K, and earlier-type M stars \citep{LUVOIR:2019, HabEx:2020_FinalReport}.

Potentially countering their observational advantages, M stars exhibit substantially different ultraviolet (UV) activity compared to stars like our Sun \citep{Stelzer:2013_UVXrayMdwarf, Shkolnik:2014, Schneider:2018, Johnstone:2021_FGKMActivity} that also changes as a function of stellar age, potentially impacting both planetary habitability and the detection and interpretation of biosignatures.  Compared to Sun-like stars, ultraviolet flux makes up a significantly larger fraction of an M star's bolometric luminosity. Additionally, the ratio of Far-UV (FUV, $\mathrm{130nm < \lambda < 170nm}$) to Near-UV (NUV, $\mathrm{170 < \lambda < 280nm}$) flux for an M star is $\approx$1 \citep[$\approx$10$^{-3}$ for Sun-like stars,][]{France:2013}.  Observations show that this high-energy (i.e.  X-ray and UV) radiation from M stars is correlated to stellar magnetic fields which are generated through rotation and convection \citep{Reiners:2022_MdwarfRotationActivity}.  As a star spins down over time, its X-ray flux remains at a ``saturated'' level until reaching a critical rotation rate, after which its X-ray flux falls as a power-law with stellar rotation rate \citep{Pizzolato:2003_XraySaturation, Preibisch:2005_XrayEvolution}.  UV flux from M stars also exhibits this saturation behavior on timescales similar to that for X-ray flux \citep{Shkolnik:2014, Schneider:2018_HAZMAT3}.  While G dwarfs like our Sun likely remain in the saturation regime for only ten to a hundred million years \citep{Ribas:2005, Tu:2015_SunlikeEvoTrack, Johnstone:2021_FGKMActivity}, early M stars (M0 -- M4) do so for a few hundred million years and late M stars (M4 -- M9) do so for up to a few billion years \citep{West:2008_AgeActivityw/SDSS, Lammer:2009, Shkolnik:2014, Schneider:2018_HAZMAT3, Johnstone:2021_FGKMActivity}.  The UV environments around M stars, especially late types, are spectrally distinct \citep[emitting less Near-UV and more Extreme-UV radiation,][]{Richey-Yowell:2023_UVEvoGaia}, and longer lived when compared to stars like our own Sun.

This distinct UV environment for M star systems can potentially change the evolution of a planet's atmospheric composition, including the chemical lifetime and abundance of biosignature molecules.  Ultraviolet light can catalyze the formation of (\ce{O2} and \ce{O3}) or directly photolyze (\ce{CH4}, \ce{H2O}, and \ce{CO2}) molecules in the atmospheres of terrestrial exoplanets, either directly through photolysis or indirectly through production of reactants/radicals \citep{Segura:2005_MdwarfBiosig}.  Photochemical models suggest that \ce{CH4} will be more abundant in the atmospheres of Earth-like planets orbiting M stars due to their relatively weak NUV flux which would otherwise produce destructive radicals from the photolysis of water and ozone \citep{Segura:2005_MdwarfBiosig, Rugheimer:2015}.  Carbon dioxide is readily photolyzed by the characteristically strong FUV flux that M stars emit to produce \ce{CO} and \ce{O}.  The direct recombination of these two is spin forbidden and slow --- chemical catalysts or collisions with a third body are needed for efficient \ce{CO2} recombination \citep{Meadows:1996, Selsis:2002} --- so M star planets with \ce{CO2} as a major constituent of their atmospheres may have a much stronger photolytic source of free oxygen radicals than G dwarf planets.  The resulting free \ce{O} can then recombine to form \ce{O2} and \ce{O3}, the latter of which is photolyzed via radiation at NUV wavelengths $\mathrm{200nm < \lambda < 300nm}$ where M stars are characteristically deficient.  Thus, \ce{O2} and \ce{O3} are thought to have stronger production and longer lifetimes on planets orbiting M stars, and several studies have explored the possibility for \ce{O2} and \ce{O3} to build up on these worlds \citep{Segura:2005_MdwarfBiosig, Segura:2007, Domagal-Goldman:2014_AbioO2O3, Tian:2014, Harman:2015_AbioOxy, Harman:2018_AbioOxyLightning, Ranjan:2022_PhotochemRunaway, Ranjan:2023_UpperAtmRunaway}.

Despite the potential for M star UV emission to alter atmospheric composition through photochemistry, the impact of stellar age-dependent changes in UV on the presence and strength of biosignatures in Earth-like planetary atmospheres has not been extensively explored.  In this study, we explore the impact of quiescent, age-dependent UV emission on the chemical compositions of these atmospheres, and the potential for detection and interpretation of biosignatures within them.  We do so by modeling high resolution (R $\mathrm{>} 10^5$) spectra of M4 and M8 stars at ages between 650 Myr and 5 Gyr as well as using previously published spectra in \citet{Peacock:2019_T1, Peacock:2020}.  We focus our study on later type M stars because, in addition to the advantages described above, their saturation phases and subsequent UV flux evolution extend well into the time period in which life is thought to have developed on Earth \citep[200 -- 800 Myr old,][]{Pearce:2018_ConstrainOoL}.  We then use these modeled stellar spectra as inputs to two different terrestrial atmospheres --- the \PIE{} ($\approx$4.5 Gyr old) and the \ARE{} ($\approx$1.5 Gyr old) --- that are simulated with a coupled photochemical-climate model to ensure that their compositions and climates are consistent with the spectral energy distribution of their host stars.  The \PIE{} is a heavily oxidized, haze-free atmosphere with a significant temperature inversion in its stratosphere due to heating via \ce{O3} absorption, while, in contrast, the \ARE{} had a much more reducing atmosphere, likely with an intermittent hydrocarbon haze \citep{Pavlov:2001_ArcheanHaze, Zerkle:2012_HazyNeoArchean}, and no stratospheric temperature inversion.


\begin{table*}[]
\caption{Stellar and planetary system parameters used in this study.  Stellar parameters are to the left of the vertical line while the planetary system parameters are to the right.  FUV and NUV fluxes for the modeled M4 \& M8 PHOENIX spectra from \citet{Peacock:2019_T1, Peacock:2020} are given at the stellar surface for all stars, and are calculated by integrating the stellar spectrum from 117 -- 178.6 nm and 177.1 -- 283.1 nm, respectively, while those for GJ 876 and TRAPPIST-1 are cited from literature.}
    \label{tab:starplanetparams}
        \centering
        \begin{tabular}{rrrrrrr|rr}
            \hline \hline
            Star & Figure & $T_\mathrm{eff}$ & log($g$) & $M_\star$ & $F_\mathrm{FUV}$ & $F_\mathrm{NUV}$ & $a$ & $P$ \\
            & Color & (K) & (cm s$^{-2}$) & ($M_\odot$) & (W/m$^2$) & (W/m$^2$) & (AU) & (solar days)\\ \hline
            GJ 876 & & 3271$\pm$157 \tablenotemark{a} & 4.87 \tablenotemark{b} & 0.32$\pm$0.03 \tablenotemark{b} & 142.9 \tablenotemark{c} & 229.7 \tablenotemark{c} & & \\
            5 Gyr M4 & \cellcolor{xkcdScarlet} & 3450 & 4.95 & 0.35 & 82.1 & 306.2 & 0.1623 & 28.9645\\
            3 Gyr M4 & \cellcolor{xkcdDandelion} & 3450 & 4.95 & 0.35 & 194.5 & 501.8 & 0.1628 & 29.0956\\
            1 Gyr M4 & \cellcolor{xkcdGrassGreen} & 3450 & 4.95 & 0.35 & 534.6 & 1530.3 & 0.1627 & 29.0739\\
            650 Myr M4 & \cellcolor{xkcdCerulean} & 3450 & 4.95 & 0.35 & 1107.0 & 1644.7 & 0.1624 & 29.0066\\
            TRAPPIST-1 & & 2566$\pm$26 \tablenotemark{d} & 5.2396$\substack{+0.0056 \\ -0.0073}$  \tablenotemark{d} & 0.0898$\pm$0.0023 \tablenotemark{e} & 49.9 \tablenotemark{f} & 185.2 \tablenotemark{f} & & \\
            5 Gyr M8 & \cellcolor{xkcdOrange} & 2559 \tablenotemark{g} & 5.28 & 0.0802 & 117.0 & 365.0 & 0.0271 & 4.0373\\
            650 Myr M8 & \cellcolor{xkcdPlum} & 2559 \tablenotemark{g} & 5.28 & 0.0802 & 772.7 & 1474.6 & 0.0242 & 3.3916\\
        \end{tabular}
        \tablerefs{(a) \citealt{Rivera2010} (b) \citealt{Stassun2019} (c) \citealt{Loyd:2016_MUSCLES3} \\ (d) \citealt{Agol:2021_T1SysParams} (e) \citealt{Mann:2019_MdwarfMassConstraint} (f) \citealt{Wilson:2021_T1MegaMUSCLES} (g) \citealt{Peacock:2019_T1}}
\end{table*}

\begin{figure*}[ht]
    \centering
    \includegraphics[width=\textwidth]{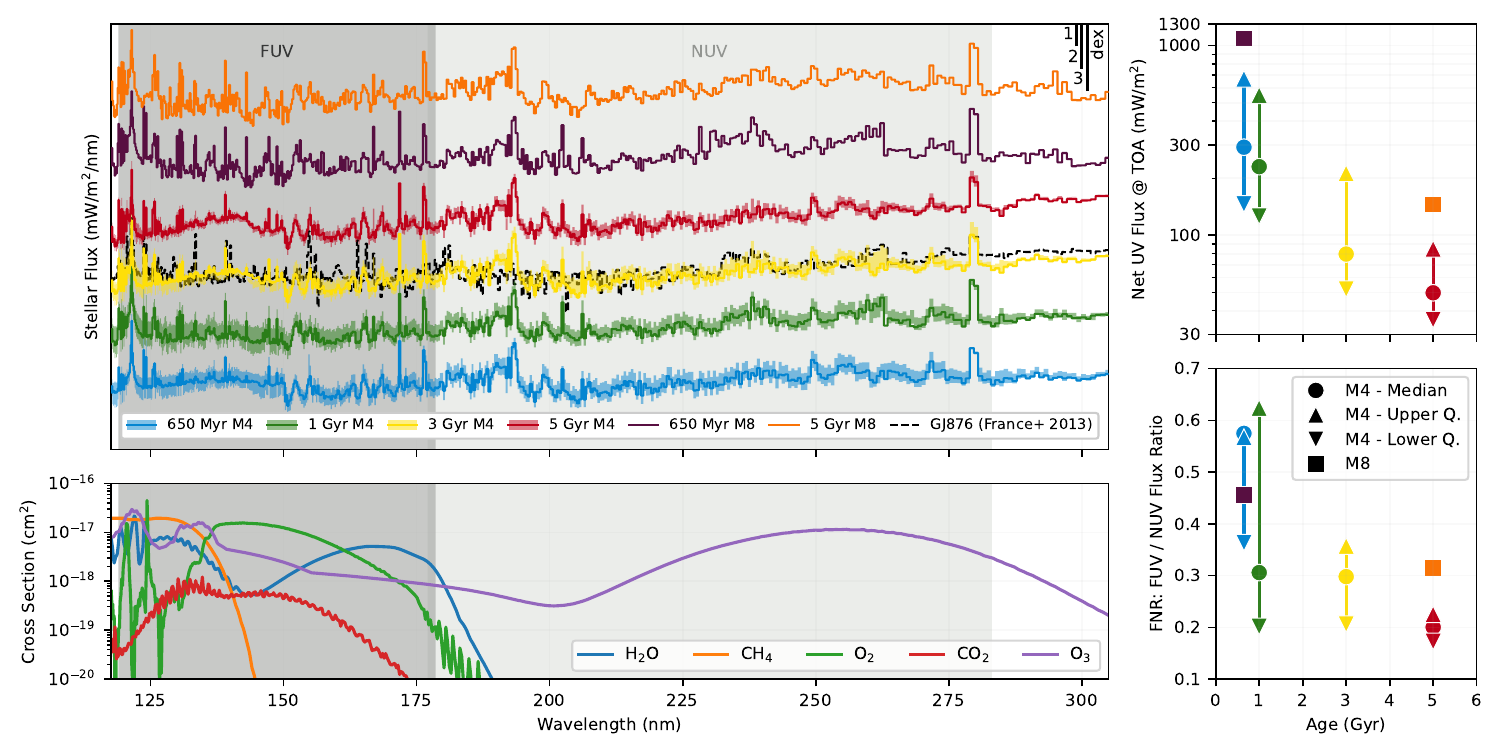}
    \caption{Modeled quiescent spectra of the M4 and M8 stars used in this study.  The top left panel shows the median spectra (solid lines, M4 and M8) as well as the interquartile range (semi-transparent envelope, M4 only) representative of M stars at each age, plus an arbitrary offset to aid with visibility.  In the absence of an absolute scale on the vertical axis, vertical lines of 1-3 dex are in the top right corner.  For comparison, we have also included the HST observations of GJ876 (M3.5V, dashed black line) from \citet{France:2016}.  The FUV band spans the 117--178.6nm range and NUV spans the 177.1--283.1nm range.  The bottom left panel depicts the absorption cross sections of molecules related to select potential biosignatures.  The bottom right panel depicts the FUV/NUV ratios (FNRs) of the median M4 (circles), upper \& lower quartile M4 (up triangle \& down triangle, respectively), and M8 (squares) spectra vs. stellar age, and shows that FNR generally decreases with stellar age.  The top right panel similarly shows that the net UV flux (integrated across the 118--300 nm range) decreases with stellar age.  The evolution of UV flux across M star age, combined with the complex UV absorption cross sections and cascading secondary chemical reaction pathways of molecules like \ce{O3} and \ce{CH4} require that detailed photochemical models be used in determining the influence of M star age on atmospheric gas abundances.}
    \label{fig:UVSpecsXsecs}
\end{figure*}

\section{Methods} \label{sec:methods}
In this section, we present our methodology for exploring the impacts of M star UV flux across stellar age on Earth-like exoplanet atmospheres.  Broadly, models and inputs are used to simulate the coupled climate-photochemical states and spectra for planets orbiting M4 and M8 dwarfs over a range of stellar ages and quiescent UV emission levels.  Our experiments focus on the potential for these atmospheres to exhibit spectral signatures and gas abundances that could be considered biosignatures, and how the range of quiescent UV emission the host star exhibits at different ages impacts the strength of those biosignatures.  We clarify that the work presented here does not consider the effect of flares on atmospheric chemistry, but rather quiescent UV emission only.

More specifically, we use a stellar atmosphere model (\S \ref{sec:methods:models:stellar}) to simulate 10 nm -- 10,000 nm spectra of M4 and M8 stars at ages between 650 Myr and 5 Gyr.  These stellar spectra are used as inputs to a photochemical model (\S \ref{sec:methods:models:photochem}) and climate model (\S \ref{sec:methods:models:climate}), which are coupled (\S \ref{sec:methods:models:coupling}) to simulate two atmospheric archetypes, the \PIE{} and the \ARE{}.  Finally, we use a radiative transfer model (\S \ref{sec:methods:models:radtrans}) to produce transit and reflectance spectra of the resulting planets.

Below, \S \ref{sec:methods:models} outlines the stellar atmosphere, photochemical, climate, and radiative transfer models, while \S \ref{sec:methods:inputs} specifies the input parameters to each of these models.

\subsection{Models} \label{sec:methods:models}

This section describes the suite of models used in this study and the experiments that led to our photochemical and climate results.  \S \ref{sec:methods:models:stellar} describes the PHOENIX stellar atmosphere code and the resulting stellar spectra produced that are used in this study.  \S \ref{sec:methods:models:photochem} details the Atmos photochemical model.  \S \ref{sec:methods:models:climate} describes the climate model, VPL Climate, used in this study.  \S \ref{sec:methods:models:coupling} details how the photochemical and climate models are coupled together to simulate the atmospheric states in \ref{sec:results}.  Finally, \S \ref{sec:methods:models:radtrans} describes SMART, the radiative transfer model we use to produce the planetary spectra presented in \S \ref{sec:results}.

\subsubsection{Stellar Atmosphere Model} \label{sec:methods:models:stellar}

To generate the quiescent stellar spectra shown in Figure \ref{fig:UVSpecsXsecs}, we used the \texttt{PHOENIX} stellar atmosphere code \citep{Hauschildt:1993, Hauschildt:2006, Baron:2007}.  PHOENIX simulates a complete model atmosphere by self-consistently computing the atmospheric structure, chemistry, and synthetic spectrum of the photosphere under the assumptions of chemical, hydrostatic, and radiative-convective equilibrium, while modeling the low-density upper atmosphere with a semi-empirical, multi-level, non-local thermodynamic equilibrium (non-LTE) code, which has atomic and molecular data suitable for the high temperatures and low densities characteristic of the upper stellar layers where UV flux originates. PHOENIX has an established history of modeling both M star photospheres \citep{Hauschildt1999,Allard2001} and M star upper atmospheres \citep{Short1998,Fuhrmeister2005,Fuhrmeister2010,Peacock:2019_T1,Peacock:2019_GJ,Peacock:2020,Hintz2019,Hintz2020}.  Recent PHOENIX spectra reproduce FUV-NIR observations of several confirmed planet-hosting M stars \citep{Peacock:2019_T1, Peacock:2019_GJ}, have been found to be the most reliable estimate of UV spectra when observations are unavailable \citep{Teal:2022_UncertaintyUV}, and have previously been used as input in studies of exoplanet photochemistry and atmospheric escape (\textit{e.g.}, \citealt{Wunderlich2020, Gialluca2021, Villarreal2021, Teal:2022_UncertaintyUV, Zieba:2023_T1cJWST, Lincowski2023_T1cAtmComp, Meadows:2023_JWSTBiosigs}).

For this paper, we use spectra representative of M4 stars at various ages from \citet{Peacock:2020} and published PHOENIX spectra of TRAPPIST-1 \citep{Peacock:2019_T1}, a M8 star with seven known terrestrial-sized exoplanets \citep{Gillon:2017_T1Discovery}.  All synthetic spectra have full wavelength coverage from 10 nm -- 10,000 nm at high-resolution (R $\mathrm{>}$ 100,000 in the 118-300 nm region), but resolutions are reduced for use in the photochemical and climate models as described in \S \ref{sec:methods:inputs:staratm}.  The HAbitable Zones and M dwarf Activity across Time (HAZMAT) spectra of the M8 star used in this work are available on MAST at \dataset[10.17909/t9-j6bz-5g89]{https://doi.org/10.17909/t9-j6bz-5g89}.

We note that \citet{Wilson:2021_T1MegaMUSCLES} published a semi-empirical spectrum of TRAPPIST-1.  This spectrum, which has been compared to \citet{Peacock:2019_T1} in \citet{Cooke:2023_T1UVO3}, used observations from HST's COS and STIS instruments covering the $100-400$ nm wavelength range.  However, the instruments were pushed to the limits of their performance with many emission lines remaining below the noise floor, and the majority of the spectrum's FUV and NUV bands are replaced with a simple polynomial fit.  Although strong, narrow emission lines likely contribute the majority of the flux in the FUV and NUV bands, weak, broad continuum emission not detected by the \citet{Wilson:2021_T1MegaMUSCLES} observations could have a strong effect on our models. Therefore, for this work we have adopted the PHOENIX spectrum of \citet{Peacock:2019_T1}, which includes a physical treatment of TRAPPIST-1's continuum emission throughout the UV range.  We discuss the potential caveats of this choice in spectral modeling in Section \ref{sec:discussion:uvchar} below.

\subsubsection{Photochemistry Model} \label{sec:methods:models:photochem}

The photochemical model used in this study, Atmos, is based on \citet{Kasting:1979} and was updated considerably in \citet{Zahnle:2006} and \citet{Lincowski:2018}.  Atmos has been validated on terrestrial planets in our own Solar System including Venus \citep{Lincowski:2018} and Earth \citep{Meadows:2023_JWSTBiosigs}.  In this study, we update the model to further include \ce{H2O} cross-sections published in \citet{Ranjan:2020_H2OXsecHab}.  This model has seen extensive use simulating many different types of atmospheres including: anoxic Archean \citep{Kharecha:2005_ArcheanAtmEco, Arney:2017_POD-Hazes}, highly oxidized (100bar \ce{O2}) \citep{Schwieterman:2016}, hazy \citep{Arney:2016_POD-Archean} and non-hazy \citep{Meadows:2018_PCb, Meadows:2023_JWSTBiosigs}, as well as atmospheres of planets in orbit around M stars \citep{Segura:2003, Segura:2005_MdwarfBiosig, Segura:2007, Domagal-Goldman:2011, Domagal-Goldman:2014_AbioO2O3, Rugheimer:2015, Lincowski2023_T1cAtmComp, Meadows:2023_JWSTBiosigs}.

\subsubsection{Climate Model} \label{sec:methods:models:climate}

To calculate pressure-temperature and water abundance profiles in this study, we use the 1D climate model VPL Climate \citep{Lincowski:2018}.  With the line-by-line radiative transfer code SMART \citep[\S \ref{sec:methods:models:radtrans},][]{Meadows:1996} as its core RT solver \citep{Robinson:2018}, VPL Climate generates layer-by-layer, line-by-line Jacobians that describe the stellar and thermal source terms, their derivatives, and the derivatives of layer reflectivity, transmissivity, and absorptivity.  These are used in a linear flux-adding framework to calculate atmospheric heating rates \citep{Robinson:2018} and step the atmosphere through time.  When that atmospheric state moves outside of the tolerance range of the Jacobians, time-stepping is halted and new Jacobians are generated.  By only re-computing the radiation field under these conditions, the atmosphere can be iterated relatively quickly while still accurately reproducing Earth-like atmospheres \citep{Meadows:2018_PCb, Meadows:2023_JWSTBiosigs}.  VPL Climate and its line-by-line radiative transfer engine, SMART, have reproduced available spectroscopic observations and climates of Earth, Venus, and Mars \citep{Robinson:2018, Lincowski:2018, Lincowski:2021_VenusSO2vsPH3, Meadows:2023_JWSTBiosigs} over wavelengths from the UV to the microwave, indicating radiative and climate physics comprehensive and accurate enough to model a diverse range of environments.  VPL Climate has previously been used to model the climates of \ARE-like worlds \citep{Meadows:2023_JWSTBiosigs}.  Since no observations of the \ARE{} exist, we have used inferred properties of these past environments from bio-geophysical proxies and model-inferred quantities.  Consequently, similar to \citet{Meadows:2023_JWSTBiosigs}, the \ARE{} cases presented here are intended to evaluate how earlier-established, less complex life could modify the environment and spectra of an exoplanet, and are not intended to be the closest possible reproduction of the \ARE.

\subsubsection{Photochemical - Climate Model Coupling} \label{sec:methods:models:coupling}

To ensure that our atmospheres are photochemically and climatically consistent with the spectral energy distributions of our M star spectra, we couple the photochemical model and climate model together.  We begin with the photochemical model's ``template'' simulations of the \PIE{} and \ARE{} atmospheres around the Sun.  For each M star-planet pair, we then point the photochemical model to the case's respective M star spectrum and input parameters described in Section \ref{sec:methods:inputs} and run the model.

Once the photochemical model has reached convergence, we input the resulting gas mixing ratio profiles to the climate model.  The climate model uses the gas abundances to calculate a new temperature-pressure structure, and condensible gas mixing ratio profiles.  Once the climate model has converged, its outputs are passed back into the photochemical model and the cycle repeats until both models reach convergence in one model run in succession.

We note that convergence criteria differ between these two models.  For the photochemical model, we define convergence as the model reaching 10$^{17}$ simulated seconds in $<$ 100 time-steps.  In contrast, the climate model will run for the user-supplied number of time-steps (10,000 in our case) and is considered converged if, for all atmospheric layers in the final state, the flux is balanced to within 1 W/m$^2$ and the heating rate is less than 10$^{-4}$ K/day \citep{Lincowski:2018}.

We clarify that the photochemical \& climate models and the results they produced are not time-dependent.  We do not simulate how atmospheres co-evolve with the stellar spectrum continuously through a given time range.  We instead compute steady-state atmospheres given a discrete stellar spectrum as input.

\subsubsection{Radiative Transfer Model} \label{sec:methods:models:radtrans}

To produce the planetary spectra we show in Figures \ref{fig:VMR_R&TSpecs_M4AgeActivity_ARE} through \ref{fig:VMR_R&TSpecs_M8AgeActivity_PIE}, we use the Spectral Mapping Atmospheric Radiative Transfer (SMART) model \citep{Meadows:1996, Crisp:1997}, a 1D, line-by-line, multiple scattering radiative transfer code.  Using novel spectral mapping techniques and the widely-used DISORT 2.0 \citep{Stamnes:1988, Stamnes:2000} code, the model efficiently simulates high resolution reflected light and emission spectra, with the capability to produce transit transmission spectra added in \citet{Robinson:2017_Transit}.  SMART ingests wavelength-dependent surface albedo, as well as gas abundances, aerosol and cloud optical depths, pressure-temperature profiles, and gas absorption coefficients.  Gas absorption coefficients are computed with the Line-by-Line Absorption Coefficient code \citep{Meadows:1996, Crisp:1997}.  SMART has been validated against observations of Mars \citep{Tinetti:2005}, Earth \citep{Robinson:2011}, and Venus \citep{Arney:2014_VenusChem}, and can model a diversity of atmospheres with multiple molecular components, as long as absorption data exists.


\begin{table*}[]
    \caption{Lower boundary conditions set in our photochemistry model.  DV refers to a deposition velocity (cm s$^{-1}$), FSF is a fixed surface flux (molecules cm$^{-2}$ s$^{-1}$), and FMR is a fixed volume mixing ratio.  Some species also have a height distribution value DISTH, indicating the number of model layers above the surface that the FSF (fixed surface flux) value is evenly divided over.}
    \label{tab:species_bcs}
    \centering
    \resizebox*{!}{0.45\textheight}
    {%
        \begin{tabular}{lccll}
        \hline \hline
        Species           & Condition & Value              & Reference                                  & Notes \\ \hline
        Both Atmospheres  &           &                    &                                            & \\ \hline
        \ce{O}            & DV        & 1                  & \citet{Domagal-Goldman:2011}               & 1 bar, \ce{N2}-dominated atmosphere \\
        \ce{H2O}          & Humidity  & 8.8$\times10^{-1}$ &                                            & \\
        \ce{H}            & DV        & 1                  & \citet{Domagal-Goldman:2011}               & 1 bar, \ce{N2}-dominated atmosphere \\
        \ce{OH}           & DV        & 1                  & \citet{Domagal-Goldman:2011}               & 1 bar, \ce{N2}-dominated atmosphere \\
        \ce{CH4}          & FSF       & 5.0$\times10^{10}$ & \citet{Seinfeld:2006}                      & Pre-industrial Earth, 150 -- 237 Tg/yr \\
        \ce{NO}           & DV        & 1.6$\times10^{-2}$ & \citet{Hauglustaine:1994_EarthChemClimate} & Continental, pre-industrial Earth \\
        \ce{NO2}          & DV        & 3.0$\times10^{-3}$ & \citet{Domagal-Goldman:2011}               & 1 bar, \ce{N2}-dominated atmosphere \\
        \ce{O3}           & DV        & 7.0$\times10^{-2}$ & \citet{Hardacre:2015_OzoneDep}             & Modern Earth $f_{dep,O3}\approx$1000 Tg/yr \\
        \ce{CO2}          & FMR       & 1.0$\times10^{-1}$ &                 & Keeps planet temperate at 0.66 \Seff, see \ref{sec:methods:inputs:atm} \\ \hline
        \PIE{} Atmosphere &           &                    &                                            & \\ \hline
        \ce{O2}           & FMR       & 2.1$\times10^{-1}$ & \citet{Seinfeld:2006}                      & Modern Earth \\
        \ce{H2}           & DV        & 2.0$\times10^{-3}$ & \citet{Steinbacher:2007_H2Flux}            & Modern Earth \\
                          & FSF       & 4.0$\times10^{8}$  & \citet{Warneck:1988_AtmChem}               & Modern Earth global volcanic flux, 200 Gg/yr \\
                          & DISTH     & 16                 &                                            & \\
        \ce{CO}           & FSF       & 1.0$\times10^{11}$ & \citet{Seinfeld:2006}                      & $\approx$ 1/3 Modern Earth's 2100 Tg/yr \\
                          & FSF       & 1.35$\times10^{9}$ & \citet{Seinfeld:2006}                      & Non-anthro, $\approx$ 1/10 Modern Earth's 70 Tg/yr \\
                          & DISTH     & 16                 &                                            & \\
        \ce{N2O}          & FSF       & 1.15$\times10^{9}$ & \citet{Seinfeld:2006}                      & Modern Earth \\ \hline
        \ARE{} Atmosphere &           &                    &                                            & \\ \hline
        \ce{O2}           & DV        & 1.0$\times10^{-4}$ & \citet{Zerkle:2012_HazyNeoArchean}         & ``Neo-\ARE'' \tablenotemark{a} \\
        \ce{H2}           & DV        & 2.0$\times10^{-3}$ & \citet{Steinbacher:2007_H2Flux}            & Modern Earth \\
                          & FSF       & 4.0$\times10^{8}$  & \citet{Warneck:1988_AtmChem}               & Modern Earth global volcanic flux, 200 Gg/yr \\
                          & DISTH     & 10                 &                                            & \\
        \ce{CO}           & DV        & 1.2$\times10^{-4}$ & \citet{Kharecha:2005_ArcheanAtmEco}        & Archean, methanogen \& acetogen-inhabited ocean\\
        \ce{CH3}          & DV        & 1                  & \citet{Domagal-Goldman:2011}               & 1 bar, \ce{N2}-dominated atmosphere \\
        \ce{C2H6}         & FSF       & 3.0$\times10^{8}$  & \citet{Nicewonger:2016_EthanePIE}          & Modern geological estimate \\
                        \end{tabular}
    }
    \tablerefs{(a) after the development of oxygenic photosynthesis (cyanobacteria) but before Great Oxidation Event}
\end{table*}

\subsection{Model Inputs} \label{sec:methods:inputs}


This section describes the key inputs for each of our models, including physical planetary and system parameters (e.g.  radius, semi-major axis), planetary surface boundary conditions, panchromatic stellar spectral energy distribution, UV--visible molecular cross sections, collision-induced absorption coefficients, rotational-vibrational line lists, and wavelength-dependent surface reflectivity (albedo).\newline\newline

\subsubsection{Stellar Atmosphere Parameters} \label{sec:methods:inputs:staratm}


We generated our quiescent, age-dependent M4 spectra using a grid of PHOENIX upper atmosphere models from \citet{Peacock:2020} that have similar stellar parameters to GJ 876, an M4 star.  The upper atmospheric models consist of ad-hoc prescribed structures for the chromosphere and transition region, attached to underlying photospheres computed for a given T$_\mathrm{eff}$, log(g), and M$_\star$.  In \citet{Peacock:2020}, 8 total grids of 46 UV spectra each were computed by systematically varying the upper atmospheric structures for 0.35 M$_\odot$ and 0.45 M$_\odot$ stars at ages of 10, 45, 120, and $\ge$200 Myr.  Stellar parameters for the models were determined from \citet[][BHAC15]{Baraffe:2015}, which indicates that T$_\mathrm{eff}$ and log(g) remain nearly constant for stars of a given mass beyond 200 Myr.  After a few hundred million years of relatively constant UV flux levels, early M type stars exhibit a decrease in NUV flux proportional to t$^{-0.85}$ and FUV flux $\propto$ t$^{-0.96}$ \citep{Shkolnik:2014}.  Using these derived relationships (equations \ref{eqn:1} and \ref{eqn:2}, below) based on FUV \& NUV photometric measurements via the Galaxy Evolution Explorer (GALEX) of stars with similar mass and age, we are able to identify synthetic spectra from the upper atmosphere grids that reproduce the expected FUV \& NUV flux for an early M star at any given age.

The left side of Table \ref{tab:starplanetparams} lists the stellar parameters for GJ 876, which are generally consistent with the 0.35 M$_\odot$ $\ge$200 Myr models from \citet{Peacock:2020}.  We note that there is large uncertainty regarding the age of GJ 876, with estimates ranging from 0.1 to 5 Gyr \citep{Rivera2005,Rivera2010,Correia2010}, and that the M stars modeled in \citet{Peacock:2020} leave the saturated regime starting at around 650 Myr.  Given this uncertainty in the M star's age and saturation limit, we select the 650 Myr and 5 Gyr models for 0.35 M$_{\odot}$ stars identified in \citet{Peacock:2020} that reproduce the median fractional FUV \& NUV flux densities from a set of 30 Hyades members (650 Myr) and 60 field stars ($\approx$5 Gyr) of spectral types M0 - M4.  We then used the relationships: 
\begin{equation} \label{eqn:1}
    F_{FUV} = 0.08 \cdot F_J \cdot t^{-0.96} 
\end{equation}
\begin{equation}\label{eqn:2}
    F_{NUV} = 0.24 \cdot F_J \cdot t^{-0.85}
\end{equation}
from \citet{Shkolnik:2014}, where $F_J$ is the Two Micron All Sky Survey (2MASS) J magnitude and $t$ is the stellar age in units of Myr, to determine the spectra with correct FUV \& NUV flux densities for 1 and 3 Gyr M4 stars.  These intermediate stellar ages were modeled to probe the photochemical and climate consequences of stellar age and activity for our Earth-like atmospheres between the two extremes discussed above.  The UV fluxes of the four models are listed in Table \ref{tab:starplanetparams}, along with those calculated from HST observations of GJ 876 \citep{France:2016} and Spitzer observations of TRAPPIST-1 \citep{Agol:2021_T1SysParams}.  We match these models to observations by performing a chi-square minimization using select emission lines (e.g.  \LyA, N V, C II, Si IV, C IV).

We note that while we model quiescent stellar spectra for 650 Myr, 1 Gyr, 3 Gyr, and 5 Gyr stars for the representative M4 stars, we only model M8 stars at 650 Myr and 5 Gyr. While M4 stars ($\approx$0.35 M$_\odot$) have ``saturation times'' --- the amount of time for which a star's XUV flux is at a maximum and is decoupled from its slowing rotation rate --- of $\approx$1 Gyr, M8 stars ($\approx$0.1 M$_\odot$) can have saturation times of $\approx$4 Gyr or more \citep{Johnstone:2021_FGKMActivity}. This means that a) the decrease in XUV flux between 650 Myr and 5 Gyr is significantly larger for an M4 star than for an M8 star (this can be seen in the upper right panel of Fig. \ref{fig:UVSpecsXsecs}) and b) 1 Gyr and 3 Gyr M8 stars are still in the saturated regime and thus nearly identical in XUV flux to a 650 Myr M8. We therefore elect to model spectra for M8 stars at 650 Myr and 5 Gyr only.

The full resolution (R $\ge$ 100,000) spectra are then binned to $R\approx$300 and $R\approx$3000 for use in our photochemical and climate models, respectively.  We confirmed that these resolutions produce nearly identical results to the full resolution input spectra while being less computationally expensive.  We do this via wavelength-averaged flux binning: for each lower-resolution wavelength bin, we use trapezoidal integration to calculate the net flux within the bin and divide the result by the bin's wavelength range.  We use a bin size of 100 wave-numbers in the photochemical model, which is equivalent to spectral resolution $R\ge$ 330 in the 118-300nm region, while the spectra used in the climate model are binned at 10 wavenumbers ($R\ge$ 3300).

\subsubsection{Modeling the Range of Stellar Activity} \label{sec:methods:inputs:activityrange}

To simplify references in the remainder of this paper to the range of quiescent UV emission exhibited by a star of a given spectral type, we will refer to the difference between the lowest and highest emission states --- across age and activity level --- as the ``considered range of UV emission'', or CRUVE, for that spectral type.  For the M4, the CRUVE can be described with the equation $$\mathrm{CRUVE_{M4}}= X_\mathrm{650 Myr M4, UQ} - X_\mathrm{5 Gyr M4, LQ}$$ and similarly for the M8 $$\mathrm{CRUVE_{M8}}= X_\mathrm{650 Myr M8} - X_\mathrm{5 Gyr M8}$$ where X is some quantity in question (net UV flux, FUV/NUV ratio, etc.) and UQ and LQ refer to the upper and lower quartile states, respectively.

Since the GALEX measurements of the many M stars used to derive equations \ref{eqn:1} and \ref{eqn:2} contain a 1--2 dex spread in measured FUV \& NUV flux at each epoch \citep{Shkolnik:2014} we model spectra that reproduce the observed median, lower quartile, and upper quartile FUV \& NUV fluxes of each M4 stellar age population bin.  Therefore, for our M4 stars, the CRUVE is the difference between the maximum UV flux model (650 Myr upper quartile) and minimum UV flux model (5 Gyr lower quartile) states.  Similarly, the CRUVE for our M8 stars is the difference between the maximum UV flux model (650 Myr) and minimum UV flux model (5 Gyr) states.  For figures that showcase the stellar spectra or atmospheric composition (e.g.  Figures \ref{fig:UVSpecsXsecs} -- \ref{fig:VMR_R&TSpecs_M8AgeActivity_PIE}), the solid colored lines reflect the median stellar activity state and, for the M4 stars and their planets only, the semi-transparent envelopes represent the interquartile ranges (IQR, 50\% range around the median) of stellar activity.

Note that as our CRUVE metric is derived from a large sample of M stars, it represents the inner range of FUV \& NUV flux densities that a star of given spectral type is likely to exhibit, not the FUV \& NUV range that an individual star will exhibit.  The IQR for the 650 Myr and 5 Gyr M4 stars were determined from GALEX measurements and bound the FUV \& NUV flux densities to within a factor of 2.  The IQRs of the 1 and 3 Gyr M4 stars are estimated by identifying models with FUV \& NUV flux densities that bound the UV emission to within a factor of $\approx$2, which is consistent with that of the observed sample.  

\subsubsection{Planetary System Parameters} \label{sec:methods:inputs:sysparams}

The planetary parameters include surface gravity, solid-body radius, rotation rate and surface albedo.  Surface gravity (g $=$ 9.8 $\mathrm{m/s^2}$ in the photochemical and climate models) and planetary radius ($R_\mathrm{p}$ = 6,371 km in both photochemical and climate models) are constant throughout all cases in this study.  Rotation rates and orbital periods are equivalent due to an assumption of synchronous rotation for these close-in M dwarf planets, and are calculated using Kepler's Third Law.  Planetary albedo is treated differently between the photochemical and climate models; a wavelength-independent surface albedo suffices for the photochemical model as photochemically active UV photons are unlikely to reach the surface and the longer wavelengths that influence climate are not involved in the photochemical reactions.  We use a wavelength-independent surface albedo of 0.25 for the \PIE{} (representative of vegetated continents) and 0.3 for the \ARE{} (representative of bare rock continents) in the photochemical model, both of which include an approximation for the effects of water clouds in the Earth’s atmosphere \citep{Kopparapu:2013}. In contrast, for the climate model, all wavelengths have the potential to modify the temperature structure of the atmosphere, especially in the visible to mid-infrared, and so the climate model’s RT solver (SMART, \S \ref{sec:methods:models:radtrans}) uses a wavelength-resolved albedo to accurately account for climatic effects that depend on the spectral shape of both the host-star’s flux and the planet’s surface \citep[e.g. ice-albedo feedback,][]{Shields:2013} and to model the reflected light spectrum of the planet \citep[e.g.][]{Madden:2020_SurfacePropsAffectClimate}.  The climate model's wavelength-resolved \PIE{} surface albedo is based on a diurnally averaged equatorial Earth view during spring equinox \citep[composite 1]{Robinson:2011}, and is comprised of 65.6\% seawater, 13.6\% grassland/brush, 11.3\% snow/ice, 5.5\% soil/desert, and 4\% conifer forest.  We use a similar albedo for the climate model's \ARE{} surface, but with the vegetated regions replaced by soil/desert such that it is comprised of 65.6\% seawater, 23.1\% soil/desert, and 11.3\% snow/ice.  These spectral surface albedos are from the USGS spectral library \citep{Clark:2007_USGS}, except for the conifer forest, which is from the ASTER spectral library \citep{Baldridge:2009_ASTERv2}.  Water clouds are explicitly modeled in the climate model \citep{Meadows:2023_JWSTBiosigs}, so no cloud approximation is included in the climate model's albedo.

System parameters include semi-major axis, orbital period, and the solar zenith angle.  The right side of Table \ref{tab:starplanetparams} details the orbital parameters we use in our model runs, and we adopt a solar zenith angle of 60$^\circ$ to approximate the global diurnal average illumination of the Earth.  The semi-major axes of our planets are calculated by integrating each spectrum over wavelength-space and using the resulting luminosity to set each planets instellation to 66\% the Earth's irradiance.  This is the reported instellation for Proxima Centauri b \citep{Anglada-Escude:2016_PCbDiscovery} and near the instellation of TRAPPIST-1e \citep{Agol:2021_T1SysParams}.  Due to the higher NIR/VIS ratio of M star spectral energy distributions compared to those of Sun-like stars, this smaller instellation prevents our terrestrial planets from entering runaway greenhouse states \citep{Kopparapu:2016}, and the resulting surface temperatures of our planets are 283.6 $\pm$ 1.6K.  We note that while the 1D climate models used here do not explicitly model the effects of tidal locking and synchronous rotation on atmospheric dynamics, 1D models have been shown to accurately model globally averaged climate states for planets with thick atmospheres, even if they are synchronously rotating \citep{Godolt:2016, Meadows:2018_PCb, Lincowski:2018}.  

\subsubsection{Atmospheric Parameters} \label{sec:methods:inputs:atm}

The modeled atmospheres presented here are comprised of many molecular species, the interactions of which are modeled via chemical reaction networks, UV--visible molecular cross-sections, and collision-induced absorption profiles.  The two atmospheres modeled in this study, the \PIE{} and \ARE{}, are described in greater detail in \citep{Meadows:2023_JWSTBiosigs}.  In the photochemical model, the \PIE{} atmospheres contain 71 species that interact via 324 reactions, with a bulk composition of 68\% \ce{N2}, 21\% \ce{O2}, and 10\% \ce{CO2}.  The \ARE{} atmospheres contain 74 species and 453 reactions, with the major gases being 10\% \ce{CO2} and 89\% \ce{N2}.  We treat lightning-induced NO$_\mathrm{x}$ chemistry in the photochemical model by scaling the modern Earth's \ce{NO} production from lightning ($\approx6\times10^8$ cm$^{-2}$s$^{-1}$) to the \ce{N2}/\ce{CO2}/\ce{O2} abundances in our atmospheres \citep[][see their Fig. 1]{Harman:2018_AbioOxyLightning}.  This induces \ce{NO} production rates of $6.1\times10^8$ cm$^{-2}$s$^{-1}$ for the \PIE{} and $2.5\times10^8$ cm$^{-2}$s$^{-1}$ for the \ARE.  Collision-induced absorption (CIA) is included in the climate model for \ce{N2}-\ce{N2} (\citet{Lafferty:1996}, as described in \citet{Schwieterman:2015b}), \ce{O2}-\ce{O2} \citep{Hermans:1999, Greenblatt:1990, Mate:1999}, and \ce{CO2}-\ce{CO2} \citep{Moore:1971, Kasting:1984, Gruszka:1997, Baranov:2004, Wordsworth:2010, Lee:2016}.  We use the HITRAN2016 line list \citep{HITRAN:2016} for IR gas absorption in our climate model and radiative transfer model.  Our atmospheres are divided into 81 layers in the climate model and 200 layers in the photochemical model, with upper atmospheric maximum altitudes of 95--104 km and minimum pressures of 38--59 nanobar. The climate-photochemcial coupling code interpolates the resulting layer grid of one model to the other's layer grid between model runs (\ref{sec:methods:models:coupling}).

Our model atmospheres conserve reduction-oxidation (redox) balance both within the atmosphere and between the atmosphere and ocean \citep{Domagal-Goldman:2014_AbioO2O3, Harman:2015_AbioOxy}.  We validate that the majority (86\%) of our models have an atmospheric redox imbalance of less than 10 ppm (one part in 100,000), while only one exhibits an imbalance of $\approx$ 40 ppm (one part in 25,000).  We also validate that the oxidizing fluxes flowing from the atmosphere into the ocean --- ranging from $6.4\times10^9$ to $4.1\times10^{10}$ cm$^{-2}$s$^{-1}$ --- are well inside the range of $F_\mathrm{floor}$ tested in \citet{Domagal-Goldman:2014_AbioO2O3}.

Our modeled atmospheres include pressure-dependent water vapor (strato-cumulus) and ice (cirrus) clouds.  The properties of these clouds and their constituent particles include pressure-dependent optical depths, phase functions or particle asymmetry parameters, extinction, scattering, and absorption efficiencies, and are pre-computed with a single scattering model for a range of particle types.  A more detailed description of the cloud treatment used in this work can be found in \citet{Meadows:2023_JWSTBiosigs}.

To support a habitable surface temperature given the reduced instellation (0.66 \Seff, \S\ref{sec:methods:inputs:sysparams}) of our modeled planets, we set the \ce{CO2} abundance of each atmosphere to 10\%, which resulted in surface temperatures of $283.6 \pm 1.6$ K.  This assumption of higher atmospheric \ce{CO2} for planets of lower instellation is not unreasonable, given that atmospheric \ce{CO2} abundance is believed to be buffered by the recycling of carbonates and silicates over geologic timescales.  This process depends on the hydrological cycle and is temperature dependent, and it has been argued that Earth would have produced higher atmospheric \ce{CO2} concentrations during cooler periods in it's history \citep{Walker:1981, Kasting:2003_EvoHabPlanet}.  Therefore, exoplanets further out in their star's HZ could also build up enough \ce{CO2} to stay temperate via this mechanism \citep{Kasting:1993}.

We did not simulate hydrocarbon hazes in our \ARE{} atmospheres as they are unlikely to form given the low P[\ce{CH4}]$/$P[\ce{CO2}] ratio of these atmospheres \citep{Trainer:2006_EarlyEarthTitanHazes}.  Hydrocarbon hazes are capable of altering an atmosphere's temperature profile and obscuring spectral features in transit transmission \citep{Arney:2016_POD-Archean}, and recent work has shown that photochemical haze formation in \ARE-like atmospheres can be sensitive --- at an observable level --- to the UV continuum of host M stars \citep{Teal:2022_UncertaintyUV}.  However, \citet{Arney:2017_POD-Hazes} showed that for P[\ce{CH4}]$/$P[\ce{CO2}] $\le 0.2$, oxygen radicals limit haze production and the resulting spectral effects in their M star + \ARE{} atmospheres.  The atmospheres in this study peak at $\ce{CH4}/\ce{CO2} = 0.07$, with most falling around $\ce{CH4}/\ce{CO2} = 0.01$, and therefore are unlikely to be subject to the photochemical, climate, and spectral effects of hydrocarbon hazes.

\begin{figure*}[ht]
    \centering
    \includegraphics[width=\textwidth]{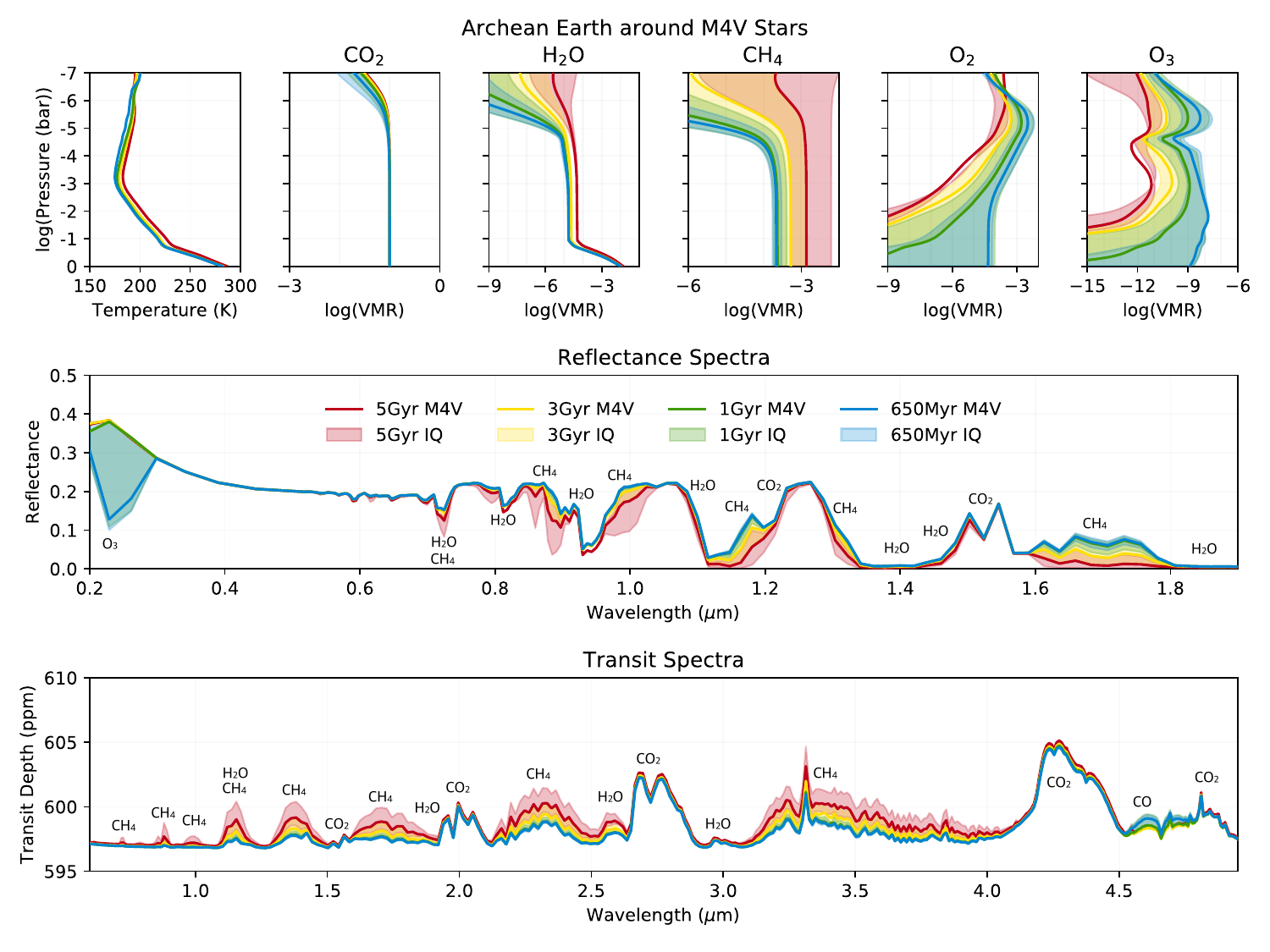}
    \caption{Atmospheric gas mixing ratios, UV-NIR reflectance spectra, and NIR transit transmission spectra of our \ARE{} analogue around an M4 star at 650 Myr (blue), 1 Gyr (green), 3 Gyr (yellow), and 5 Gyr (red).  For each panel, planets irradiated by median spectra at each age are represented with solid lines, while each ages inner quartile (IQ) range is represented by the semi-transparent envelope.  The top row of panels shows mixing ratios of five atmospheric gases relevant to climate, biosignature detection, and biosignature interpretation.  The middle panel shows reflectivity spectra --- the planet's reflected flux divided by incident stellar flux --- at an orbital phase angle of 60$^\circ$ for each planetary atmosphere binned to the resolution of the proposed LUVOIR-B concept mission \citep[$R_\mathrm{UV}\approx$ 7, $R_\mathrm{vis}=$ 140, $R_\mathrm{NIR}=$ 70,][]{LUVOIR:2019}.}  The bottom panel depicts the transit transmission spectrum of these atmospheres binned to the resolution of JWST's NIRSpec Prism instrument.  Key molecular features in the spectra are labeled. The 0.26\um{} \ce{O3} Hartley band reflectance feature is deeper at 650 Myr and 1 Gyr due to \ce{O3} buildup from extensive \ce{CO2} photolysis, but disappears at later stellar ages.
    \label{fig:VMR_R&TSpecs_M4AgeActivity_ARE}
\end{figure*}

\begin{figure*}[ht]
    \centering
    \includegraphics[width=\textwidth]{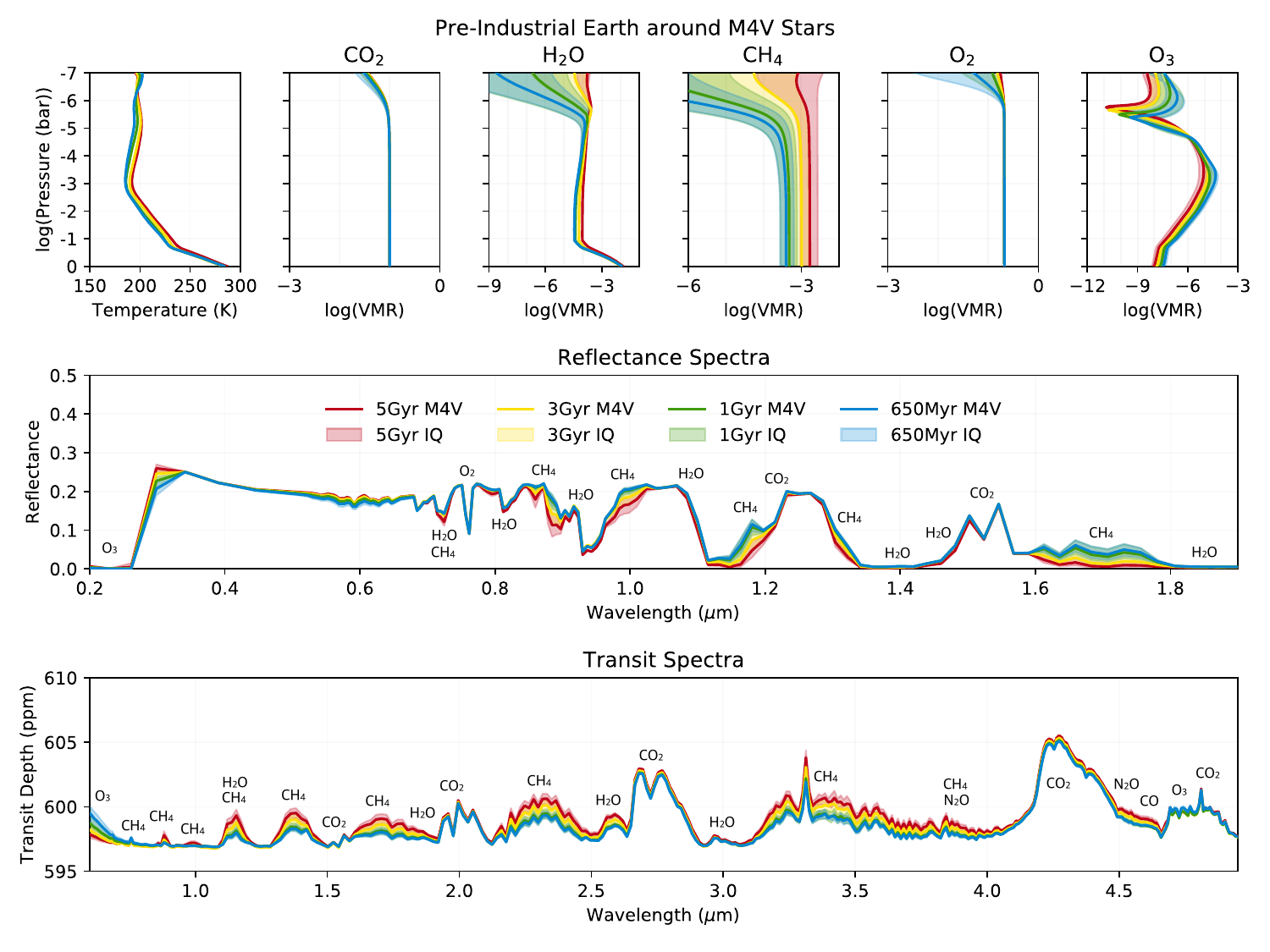}
    \caption{The same figure as Fig. \ref{fig:VMR_R&TSpecs_M4AgeActivity_ARE}, but with the \PIE{} atmospheres.  In comparison with the previous figure, the effects of the M4 considered range of UV emission (CRUVE, \S\ref{sec:methods:inputs:activityrange}) on the magnitude of changes to species abundance (top row of panels) and spectral feature strength (bottom two panels) are slightly weaker for the \PIE{} than for the \ARE.  This is especially prominent for the 0.26\um{} \ce{O3} Hartley band reflectance feature: while now much deeper at all host-star ages than in the M4 + \ARE{} (Fig. \ref{fig:VMR_R&TSpecs_M4AgeActivity_ARE}, the weakening of the feature with increasing host-star age now primarily occurs in the wings of the band.}
    \label{fig:VMR_R&TSpecs_M4AgeActivity_PIE}
\end{figure*}

\begin{figure*}[ht]
    \centering
    \includegraphics[width=\textwidth]{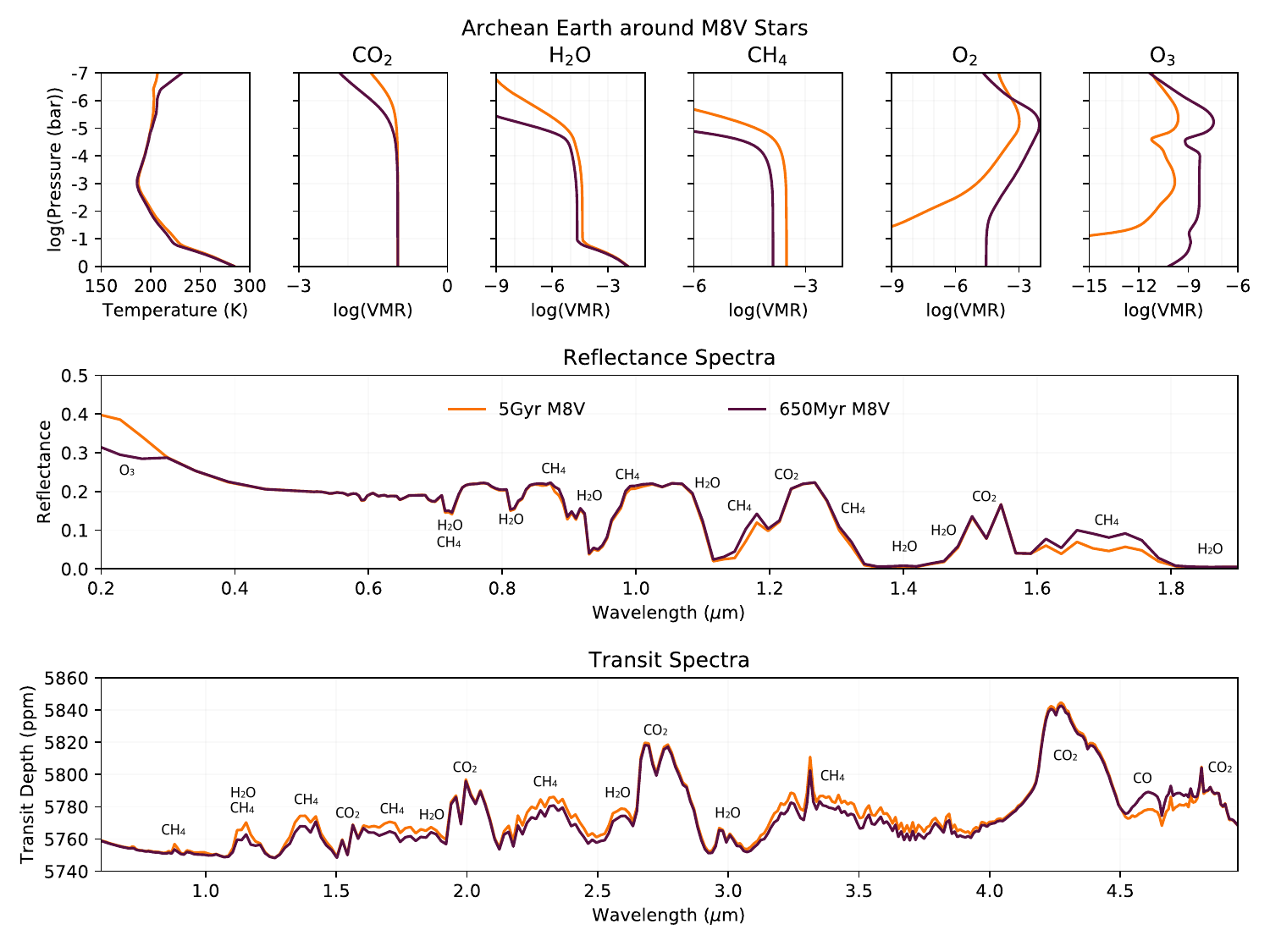}
    \caption{Atmospheric gas volume mixing ratios (VMR), UV-NIR reflectance spectra, and NIR transit transmission spectra of our \ARE{} analogue around an M8 star at 650 Myr (purple) and 5 Gyr (orange).  The top row shows mixing ratios of five critical atmospheric gases with regards to climate and biosignature detection.  The middle panel shows reflectivity spectra --- the planet's reflected flux divided by incident stellar flux --- at an orbital phase angle of 60$^\circ$ for each planetary atmosphere binned to the resolution of the proposed LUVOIR-B concept mission \citep[$R_\mathrm{UV}\approx$ 7, $R_\mathrm{vis}=$ 140, $R_\mathrm{NIR}=$ 70,][]{LUVOIR:2019}.  The bottom panel depicts the transit transmission spectra binned to the resolution of} JWST's NIRSpec Prism instrument.  Key molecular features in the spectra are labeled.  Similar to Fig. \ref{fig:VMR_R&TSpecs_M4AgeActivity_ARE}, the 0.26\um{} \ce{O3} Hartley band reflectance feature is slightly deeper at 650 Myr due to \ce{O3} buildup from extensive \ce{CO2} photolysis, but disappears at later ages.
    \label{fig:VMR_R&TSpecs_M8AgeActivity_ARE}
\end{figure*}

\begin{figure*}[ht]
    \centering
    \includegraphics[width=\textwidth]{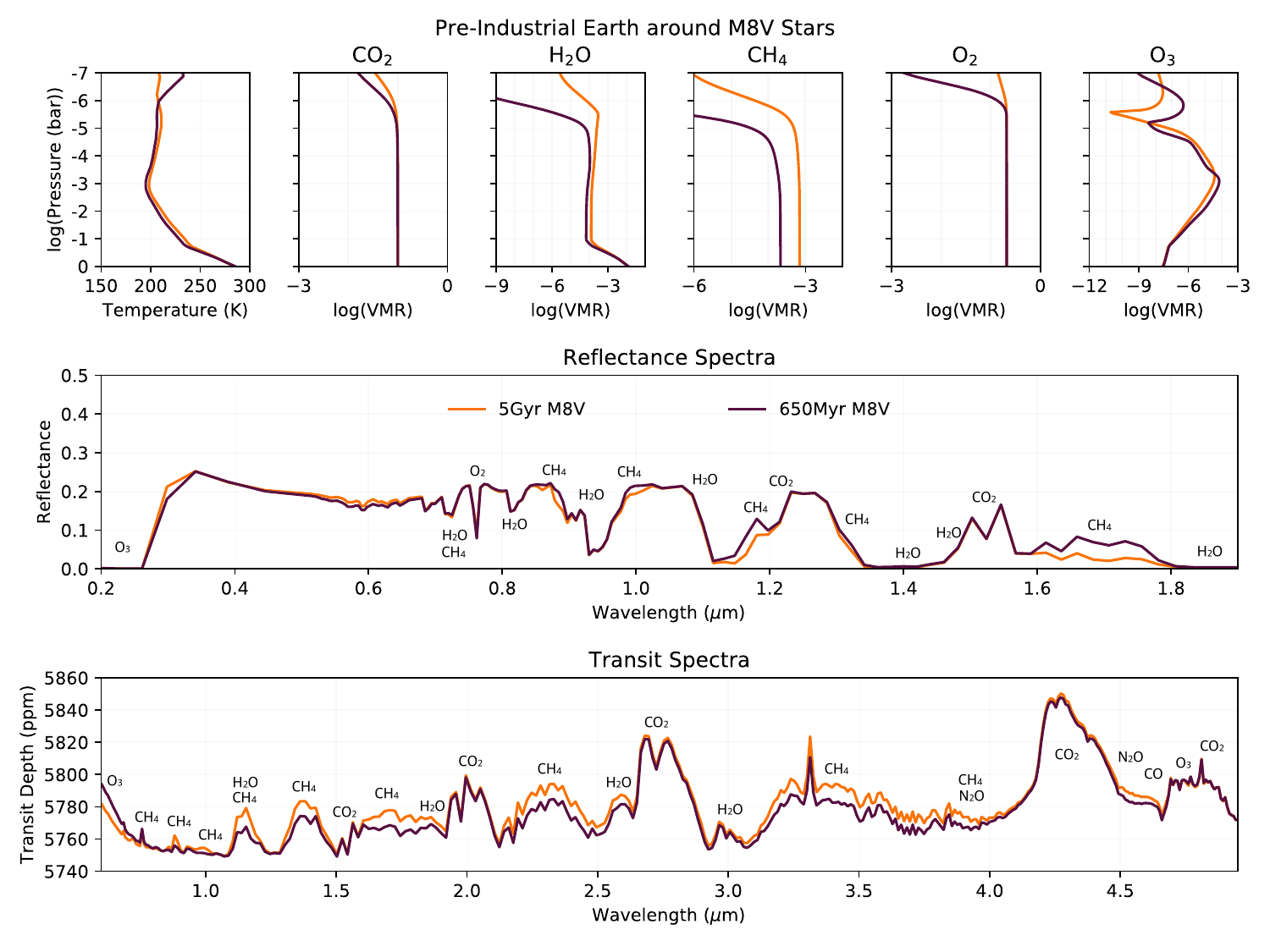}
    \caption{The same figure as Fig. \ref{fig:VMR_R&TSpecs_M8AgeActivity_ARE}, but with the \PIE{} atmospheres.  In comparison with the previous figure, the magnitude of changes to species abundance (top row of panels) and spectral feature strength (bottom two panels) due to the M8 considered range of UV emission (CRUVE, \S\ref{sec:methods:inputs:activityrange}) are slightly weaker for the \PIE{} than for the \ARE.  This is especially prominent for the 0.26\um{} \ce{O3} Hartley band reflectance feature: while now much deeper at all host-star ages than in the M4 + \ARE{} (Fig. \ref{fig:VMR_R&TSpecs_M4AgeActivity_ARE}, the weakening of the feature with increasing host-star age now primarily occurs in the wings of the band.}
    \label{fig:VMR_R&TSpecs_M8AgeActivity_PIE}
\end{figure*}

\section{Results} \label{sec:results}

Here we show the simulated impact of UV radiation from M stars at our sampled ages on planetary atmospheric composition and spectra, and describe the underlying photochemical processes.  Sections \ref{sec:results:MAPSF_M4} and \ref{sec:results:MAPSF_M8} describe how the abundances of molecules and their spectral features change with stellar age around M4 and M8 stars, respectively, while Section \ref{sec:results:FNRvsNetUV} shows how the abundances of two canonical biosignature molecules, \ce{O3} and \ce{CH4}, are sensitive to the net UV flux and FUV/NUV ratio (FNR) of the host star.

As the spectral features in transit transmission and reflected light spectroscopy differ in appearance, we clarify that an increase in the abundance of a molecule will increase the effective cross-section of the planets atmosphere and result in an increase in the transit spectroscopy signal from the planet (higher ``hills'').  In reflectance spectroscopy, that same increase in molecular abundance will further attenuate the light on its incoming and outgoing paths through the planets atmosphere, decreasing the reflectivity of the planet (deeper ``valleys'').




\subsection{Molecular Abundance Profiles and Spectral Features of Planets Orbiting M4 Stars} \label{sec:results:MAPSF_M4}

Our simulated atmospheric abundance profiles as well as transit and reflectance spectra for planets orbiting M4 stars at ages from 650 Myr to 5 Gyr are shown in Figure \ref{fig:VMR_R&TSpecs_M4AgeActivity_ARE} for the \ARE{} and Figure \ref{fig:VMR_R&TSpecs_M4AgeActivity_PIE} for the \PIE.  We find that abundances of typical biosignature molecules like \ce{O2}, \ce{O3}, and \ce{CH4} can vary by several dex across the ``considered range of UV emission'' (CRUVE $=$ young to old, blue to red, \S\ref{sec:methods:inputs:activityrange}) of M4 stars, causing features in reflectance spectra features to change by up to 0.15 in reflectivity and transit spectra features to change by up to 60\% (3.1 parts per million (ppm)).

Over the M4 CRUVE, the \ARE{} \ce{CH4} column depth increases by 1.6 dex and the \ce{O3} column depth decreases by 5.4 dex.  The high \ce{O3} concentrations in the 650 Myr median and upper quartile and 1 Gyr upper quartile cases produce a strong 0.26 \um{} Hartley band reflectance spectrum feature, which can be seen on the left side of the middle panel in Fig. \ref{fig:VMR_R&TSpecs_M4AgeActivity_ARE}.  The \ce{CH4} surface abundance rises from 173 ppm to 6234 ppm.  This 36x increase in abundance produces a 2 ppm (68\%) increase in the 1.7 \um{} \ce{CH4} transmission feature and a 0.07 (90\%) deepening in reflectivity in the 1.7 \um{} \ce{CH4} reflectance feature.  In addition, the 1.0 \um{} \ce{CH4} reflectance feature deepens by 0.09 (45\%) in reflectivity.  For \ce{H2O}, the column depth increases by 55\%, but many of water's spectral features are overlapping with methane's in the case of transmission.  While water does have strong reflectance features at 0.93\um{} and 1.9\um{}, these features vary weakly as a function of stellar host age (0.02 deepening in reflectivity at 0.93\um{} in reflectance).  The incoming stellar UV radiation also affects the abundances of \ce{O2} and \ce{O3} in the \ARE{} atmospheres and enhances abiotic production of these gases when the host star has higher UV emission.  The surface abundance of (abiotically produced) \ce{O2} reaches on the order of 50-100 ppm around the 650 Myr median and upper quartile and 1 Gyr upper quartile, but is near-zero ($\mathrm{p[O_2] \ll 10^{-11}}$) in the remaining cases.  Despite the 50-100 ppm \ce{O2} abundances in the young host M4 cases, this is very small compared to Earth's 21\% \ce{O2}, and the young M4 abundance is not enough to produce significant spectral features in either the transmission or reflected light spectra.  In contrast, the \ce{O3} column depth decreases by over 5 dex, from 1.1$\times$10$^{17}$ to just 4.5$\times$10$^{11}$, over the M4 CRUVE.  As a result, for the \ARE{} irradiated by the 650 Myr upper quartile, 650 Myr median, and 1 Gyr upper quartile M4 stars, the region of the Hartley band from 0.2--0.3 \um{} displays a deep \ce{O3} absorption that is half the reflectance seen for planets orbiting the older, less emissive M4 stars.

The CRUVE of our M4 stars is less effective at altering the chemistry of the \PIE{} atmospheres.  Figure \ref{fig:VMR_R&TSpecs_M4AgeActivity_PIE} shows that the \ce{CH4} surface abundance increases by one dex, the \ce{O3} column depth drops by one dex, and the subsequent changes to the planetary spectra are more muted than they are in the \ARE.  The \ce{CH4} surface abundance increases by one dex, causing a 1.3 ppm (41\%) increase and 0.05 reflectivity (85\%) decrease in the 1.7 \um{} transit and reflectance features, respectively.  The \ce{H2O} column depth increases by 27\%, but its spectral features are still weakly changed or overlapping with methane features that experience larger changes.  The surface mixing ratio of \ce{O2} of 21\% (\S \ref{sec:methods:inputs:atm}) is maintained throughout most of the atmospheric column except at very low pressures ($P\le10^{-6}$ bar) where \ce{O2} destruction is stronger for the younger, higher UV output stars.  In these \PIE{} atmospheres, there is enough oxygen to produce small features at 0.76\um{} and 1.27\um{} in reflectance, but they are poorly resolved at the resolution of LUVOIR B as given by the Planetary Spectrum Generator \citep{Villanueva:2018_PSG}.  The column depth of \ce{O3} falls by one dex with increasing stellar age, and while the wing of the Hartley band ($\approx$0.3\um{}) rises by 0.05 (50\%) in reflectivity over the M4 CRUVE, the majority of the band is strongly saturated for all host star ages.

\subsection{Molecular Abundance Profiles and Spectral Features of Planets Orbiting M8 Stars} \label{sec:results:MAPSF_M8}

Figure \ref{fig:VMR_R&TSpecs_M8AgeActivity_ARE} shows how the \ARE{} responds to instellation from M8 stars across their considered range of UV emission.  Across the M8 CRUVE (young to old, purple to orange, see Section \ref{sec:methods:inputs:activityrange}), the \ce{CH4} surface abundance increases by 0.4 dex, and the \ce{O3} column depth decreases by 3.1 dex.  For the 650 Myr M8 star, the \ce{O3} abundance produces a weaker Hartley band feature than exhibted by the M4 planets (see Fig. \ref{fig:VMR_R&TSpecs_M4AgeActivity_ARE}).  The 0.4 dex increase in surface abundance of \ce{CH4} over the CRUVE produces an increase of 5.1 ppm (24\%) in the 1.7\um{} \ce{CH4} transmission feature and a deepening of 0.03 (34\%) in reflectivity of the 1.7\um{} reflectance feature.  \ce{H2O}'s column depth increases 8\% but its spectral features are still weakly varying or overlapping with strong \ce{CH4} features.  The surface abundance of \ce{O2} reaches 30 ppm around the 650 Myr M8, which is still too low to induce absorption features that are discernible by eye in the spectra.  The column depth of \ce{O3} decreases by 3.1 dex over the M8 CRUVE, causing the Hartley band (0.26\um{}) absorption feature to weaken by 0.08 (27\%) in reflectance.

Similar to the M4 cases, the M8 CRUVE has a smaller impact on the \PIE{} than the \ARE.  Figure \ref{fig:VMR_R&TSpecs_M8AgeActivity_PIE} shows the \PIE's \ce{CH4} surface abundance increasing by 0.5 dex and the \ce{O3} column depth decreasing by only 0.3 dex across the M8 CRUVE.  The \ce{CH4} surface abundance increase of 0.5 dex causes a 8.1 ppm (32\%) increase and 0.03 deepening (53\%) of the 1.7 \um{} transit and 1.7 \um{} reflectance features, respectively.  The \ce{H2O} column depth only rises by 1\%.  \ce{O2} is 21\% throughout the atmosphere (\S \ref{sec:methods:inputs:atm}) except for high in the atmosphere ($\mathrm{\le10^{-6}}$ bar).  Here, its abundance drops due to photodissociation from the M8 UV emission, an effect which is considerably stronger around the 650 Myr M8 host star.  The 0.76\um{} and 1.27\um{} \ce{O2} mentioned above are still indiscernible by eye.  The column depth of \ce{O3} decreases by only 0.28 dex, and so the \ce{O3} abundance is sufficiently high in both cases that the Hartley band is still strongly saturated.  The wing of this band ($\approx$0.3\um{}) increases by 0.02 (18\%) in reflectivity between the two cases.

\begin{figure}[ht]
    \centering
    \includegraphics[width=\linewidth]{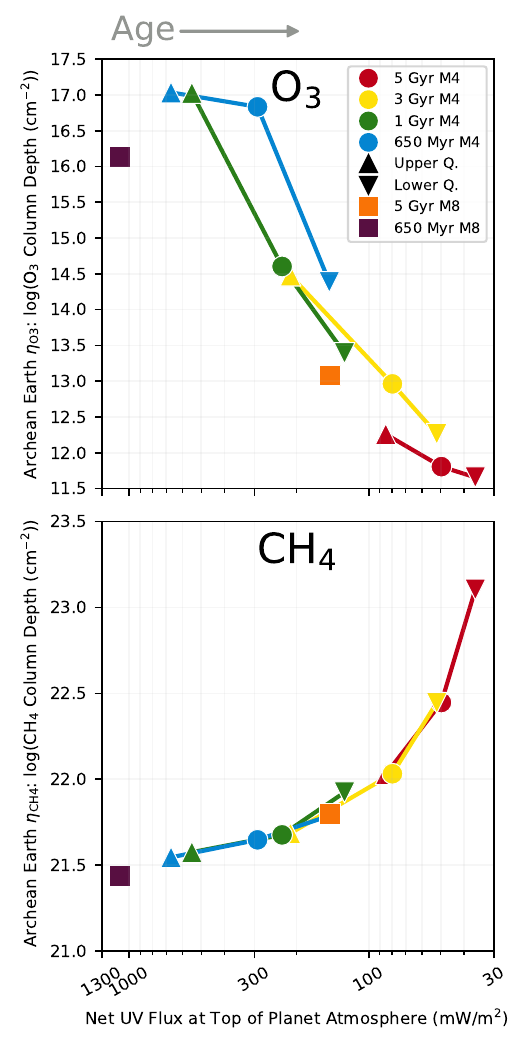}
    \caption{Column depths of ozone (top panel) and methane (bottom panel) in the \ARE{} as a function of the incident net UV flux (118--300 nm) at the top of the atmosphere.  The horizontal axis is reversed (incident net UV flux decreases to the right) so that stellar age increases to the right, as indicated by the grey arrow.  Median (circle), upper quartile (up triangle), and lower quartile (down triangle) UV emission cases for each M4 age are connected by lines to guide the eye.  M8 cases have no quartiles and are shown as squares.  In general, ozone column depth decreases while methane's column depth increases as net UV flux falls.  The \ARE's ozone column depth is most sensitive to net UV flux between 300--100 mW m$^{-2}$, decreasing by $\approx$4--5 dex. Methane's column depth is most sensitive from 100--30 mW m$^{-2}$ net UV flux, where it increases by about one dex.}
    \label{fig:NetUVvsMolCD_ARE}
\end{figure}




\subsection{Photochemical Effects of Net UV Flux and Far-UV/Near-UV Ratio} \label{sec:results:FNRvsNetUV}

As Figure \ref{fig:UVSpecsXsecs} shows, the net UV flux and FUV to NUV ratio (FNR) of our M stars generally decrease with age.  Figure \ref{fig:NetUVvsMolCD_ARE} and Figure \ref{fig:FNRvsMolCD_ARE} illustrate the relative importance of these two variables in determining the chemistry of our \ARE{} atmospheres.  We draw the reader's attention to the decreasing horizontal axis in both figures so that stellar age generally increases to the right.  Fig. \ref{fig:NetUVvsMolCD_ARE} shows $\eta_{\mathrm{O3}}$ and $\eta_{\mathrm{CH4}}$, the logarithmic column depths of \ce{O3} and \ce{CH4} respectively, in the \ARE{} as a function of the net UV flux (integrated from 118--300 nm) for each spectrum.  Fig. \ref{fig:FNRvsMolCD_ARE} is similar, but shows the column depths of these molecules when that case's UV spectrum has been scaled so that its net UV flux is equal to the 1 Gyr median case (236.6 mW m$^{-2}$) while still retaining its original FNR.  Essentially, Fig. \ref{fig:NetUVvsMolCD_ARE} shows the \ARE{} under the effects of both net UV flux and FNR, while Fig. \ref{fig:FNRvsMolCD_ARE} isolates the effect of the FNR.

First, we note how net UV flux and FNR change with stellar age.  Both Figure \ref{fig:NetUVvsMolCD_ARE} and Figure \ref{fig:FNRvsMolCD_ARE} show that younger, more active stars (purple, blue, and some green markers) tend to have higher net UV flux as well as higher FNR than their older counterparts (yellow, orange, and red markers).  The upper quartile states (up triangles) of the M4 cases have higher net UV fluxes and typically higher FNRs than their respective median states, while the lower quartile states (down triangles) have lower net UV fluxes and typically lower FNRs. We also draw the reader's attention to the scale of the vertical axes in Fig. \ref{fig:NetUVvsMolCD_ARE} and Fig. \ref{fig:FNRvsMolCD_ARE}, where $\eta_{\mathrm{O3}}$ (top panel of both figures) covers six dex, while $\eta_{\mathrm{CH4}}$ (bottom panel of both figures) covers just 2.5 dex.

\subsubsection{Ozone} \label{sec:results:FNRvsNetUV:o3}

We find that the ozone column depth in our \ARE{} atmospheres decreases with stellar age as both the host star's net UV flux and FNR decrease.  Figure \ref{fig:NetUVvsMolCD_ARE} shows that when our host M stars are young (650 Myr to 1 Gyr) and the \ARE{} atmospheres are receiving more than $\approx$300 mW m$^{-2}$ of UV flux, \ce{O3} column depth is at its highest ($\eta_{\mathrm{O3}}\gtrsim$ 16).  The \ce{O3} column depth becomes much more sensitive when net UV flux falls to $\approx$200 mW m$^{-2}$, where a $\approx$30\% decrease in the net UV flux leads to a $\approx$3 dex drop in the ozone column depth, from $\eta_{\mathrm{O3}}\approx$ 16.5 to $\eta_{\mathrm{O3}}\approx$ 13.5.  For example, the \ARE{} atmosphere has a relatively high ozone column depth of $\eta_{\mathrm{O3}}=$ 16.8 when irradiated at 291 mW m$^{-2}$ from the 650 Myr M4 (blue circle), but this drops  to $\eta_{\mathrm{O3}}=$ 14.6 around the 1 Gyr M4 (green circle) at 230 mW m$^{-2}$, a two dex decrease in $\eta_{\mathrm{O3}}$ after a mere 60 mW m$^{-2}$ reduction in net UV flux). 

\begin{figure}[ht]
    \centering
    \includegraphics[width=0.925\linewidth]{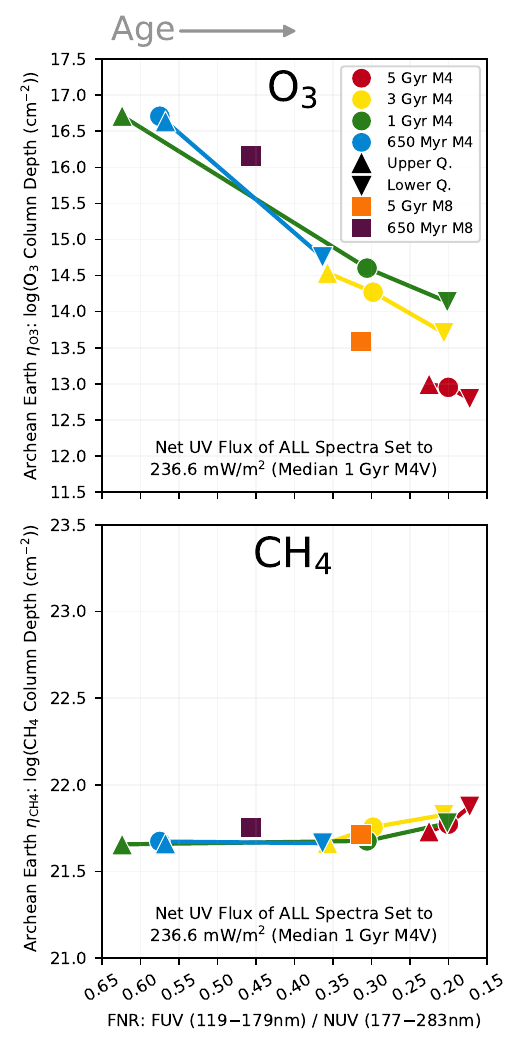}
    \caption{To isolate the photochemical effect of the FNR, this figure is similar to Fig. \ref{fig:NetUVvsMolCD_ARE} but the incident stellar spectra have been scaled between 118--300 nm so that they all possess the same net UV flux as the 1 Gyr median case (236.6 mW m$^{-2}$) while their FNRs (as seen in the bottom right panel of Fig. \ref{fig:UVSpecsXsecs}) are maintained.  The horizontal axis is reversed (FNR decreases to the right) so that stellar age increases to the right, as indicated by the grey arrow.  Symbols are the same as Fig. \ref{fig:NetUVvsMolCD_ARE}.  In general, the ozone column depth shows a significant positive correlation with FNR, while the methane column depth is minimally impacted by FNR.  The correlation between FNR and the \ARE's \ce{O3} column depth exhibits less scatter than in Fig \ref{fig:NetUVvsMolCD_ARE}, and decreases by $\approx$3 dex when FNR falls from 0.6 to 0.2, illustrating how ozone's abundance is impacted by both net UV flux and FNR.  Over that same decrease in FNR, the methane column depth increases by only 0.2 dex, suggesting that the abundance of methane in our \ARE{} atmospheres is primarily sensitive to the net UV flux (see Fig. \ref{fig:NetUVvsMolCD_ARE}) and relatively insensitive to the FNR.}
    \label{fig:FNRvsMolCD_ARE}
\end{figure}

Where the net UV flux nears $\approx$200 mW m$^{-2}$, FNR becomes an important secondary effect in determining the abundance of \ce{O3}.  Fig. \ref{fig:FNRvsMolCD_ARE} shows that, when each planet is receiving $\approx$240 mW m$^{-2}$ of UV flux but the FNR of each spectrum remains the same as the bottom right panel of Fig. \ref{fig:UVSpecsXsecs} (\S \ref{sec:results:FNRvsNetUV}), high FNRs ($\gtrsim$0.55) produce ozone-rich ($\eta_{\mathrm{O3}}\approx$ 16.5) \ARE s, while lower FNR ($\approx$0.2) produces $\eta_{\mathrm{O3}}$ from 14.1 to 13.0.  This is primarily because FUV photons drive extensive \ce{CO2} photolysis that releases free oxygen radicals which form \ce{O3}, while NUV flux directly photolyzes \ce{O3} (see bottom panel of Fig. \ref{fig:UVSpecsXsecs}.)  Interestingly, Fig. \ref{fig:NetUVvsMolCD_ARE} shows that while the 650 Myr M8 star (purple square) is more UV-luminous in the HZ than its M4 equivalent (blue circle), it produces less (0.5 dex) ozone in the \ARE; Fig. \ref{fig:FNRvsMolCD_ARE} shows that the FNR of the 650 Myr M8 star is about 80\% that of the 650 Myr M4.  Our results therefore suggest that both net UV flux and FNR play a key role in whether \ce{O2}/\ce{O3} buildup occurs in these anoxic atmospheres. 



\subsubsection{Methane} \label{sec:results:FNRvsNetUV:ch4}

We find that the increase in methane as the host M star ages is significantly more correlated with the net UV flux than the FNR of the host star.  Figure \ref{fig:NetUVvsMolCD_ARE} shows that, when net UV flux is above $\approx$130 mW m$^{-2}$, methane's column depth is weakly anti-correlated with net UV flux, increasing by 0.5 dex between $\eta_{\mathrm{CH4}}=$ 21.4 at the 1093 mW m$^{-2}$ of the 650 Myr M8 star (purple square) to $\eta_{\mathrm{CH4}}=$ 21.9 at the 126 mW m$^{-2}$ of the 1 Gyr lower quartile M4 star (green down triangle).  However, as net UV flux falls below $\approx$130 mW m$^{-2}$, methane's column depth becomes increasingly anti-correlated with net UV flux, increasing by 1.2 dex to $\eta_{\mathrm{CH4}}=$ 23.1 at the 36 mW m$^{-2}$ net UV flux of the 5 Gyr lower quartile M4 star (red down triangle).  The strong correlation of methane column depth with net UV flux, and its much weaker correlation with FNR, are attributable to the main sources of \ce{CH4} destruction in our models.  In the tenuous upper atmosphere (P $\lesssim$ 100 $\mu$bar), methane is destroyed primarily by direct photolysis via FUV radiation ($\lambda\lesssim$ 155 nm).  In the dense lower atmosphere (P $\gtrsim$ 100 $\mu$bar), methane is destroyed primarily by reaction with OH sourced from water photolysis.  While water is photolyzed at both FUV and NUV wavelengths ($\lambda\lesssim$ 240 nm), in the lower atmosphere this photolysis is primarily driven by NUV flux as most FUV radiation is absorbed in the upper atmosphere \citep{Harman:2015_AbioOxy}.  Thus, methane’s atmospheric lifetime in our models is largely a function of the incident NUV flux, rather than FNR.


In contrast, Figure \ref{fig:FNRvsMolCD_ARE} shows that the decreasing FNR of our host M stars with age minimally increases the column depth of \ce{CH4}.  In general, when each planet is receiving $\approx$240 mW m$^{-2}$ of UV flux but the FNR of each spectrum remains the same as the bottom right panel of Fig. \ref{fig:UVSpecsXsecs} (\S \ref{sec:results:FNRvsNetUV}), $\eta_{\mathrm{CH4}}$ increases by just 0.2 dex as FNR falls from $\approx$0.6 to $\approx$0.2.  The impact of FNR on $\eta_{\mathrm{CH4}}$ is much smaller than the 0.5--1.5 dex changes imparted from varying levels of net UV flux (\S \ref{sec:results:MAPSF_M4}), and so we expect the effects of FNR on methane's abundance to also be well below the noise floor of JWST around mid- and even late-type M stars.\newline\newline

\section{Discussion} \label{sec:discussion}

We have modeled the photochemical and climatic impact of M star age-dependent UV flux on Earth-like atmospheres, and find that abiotic ozone may be produced in \ARE-type atmospheres around young M stars, with up to 5 dex more ozone in a 10\$ \ce{CO2} \ARE-like atmosphere around 650 Myr M4 than a 5 Gyr old M4 star.  We also find that Earth-like planets around old M stars have longer atmospheric lifetimes for \ce{CH4}, allowing accumulation of up to 36 times as much methane in their atmospheres than planets with similar surface fluxes of \ce{CH4} around young M stars.  Additionally, we find that the methane abundance of our \ARE{} atmospheres are strongly correlated with the net UV flux coming from the host star and that the \ARE's ozone abundance has complex dependencies on both the net UV flux and the ratio of FUV to NUV flux coming from the host star.

The strong ozone Hartley-band features described above --- resulting from abiotic \ce{O2} and \ce{O3} buildup via \ce{CO2} photolysis --- is unlikely to be detectable with current telescopes. However, it may constitute a false positive biosignature for oxygenic photosynthesis in observations of Earth-like exoplanets with future UV direct imaging observatories like the Habitable Worlds Observatory.  Additionally, while the M4 planets considered here see large changes in methane abundance dependent on the host star's age and UV emissivity, the resulting atmospheric transmission features are only of the order of a few ppm, and so even these large changes in abundance are unlikely to be detectable with JWST. 

In the following sections, we discuss our results.  In \S \ref{sec:discussion:shielding}, we discuss how photochemical shielding leads to the reduced impact of stellar-age-dependent UV emission on the \PIE{} vs. the \ARE.  Section \ref{sec:discussion:o2o3buildup} discusses how \ce{O2} and \ce{O3} build up abiotically in our \ARE{} atmospheres, and how this abiotic production compares to those of previous studies.  Section \ref{sec:discussion:robustO2O3} discusses how the \ce{O2} and \ce{O3} buildup behavior presented here is robust to unphysical and model-driven phenomena.  In Section \ref{sec:discussion:observing}, we discuss which atmospheric composition changes are likely or unlikely to be observable with current and future observatories.  Section \ref{sec:discussion:biosigs} discusses how our abiotic \ce{O2}/\ce{O3} buildup compares to and may be distinguished from previously suggested \ce{O2}/\ce{O3} biosignatures.  Finally, Section \ref{sec:discussion:uvchar} discusses how characterization of M-dwarf UV flux and activity will be critical in interpreting the composition of exoplanet atmospheres in the future.

\subsection{Photochemical Shielding} \label{sec:discussion:shielding}

For both the M4 and M8 spectral types, we found that our \PIE{} atmospheres were less sensitive to age-dependent M star UV flux than our \ARE{} atmospheres (c.f. Figures \ref{fig:VMR_R&TSpecs_M4AgeActivity_ARE}--\ref{fig:VMR_R&TSpecs_M8AgeActivity_PIE}, which is likely due to differences in the strength of photochemically mediated UV-shielding.  The \PIE{} has abundant (21\%) \ce{O2}, which has a strong cross section in the FUV (see lower left panel of Fig. \ref{fig:UVSpecsXsecs}).  The short-wave ($\lambda < $ 175nm) radiation incident on the \PIE{} is strongly absorbed by \ce{O2} high in the atmosphere, which a) shields molecules in the deep atmosphere from photodissociation, and b) produces substantial \ce{O3} via \eqref{CM:R1} and \eqref{CM:R2} of the Chapman mechanism \citep{Chapman:1930_Ozone}: 
\begin{align}
    \ce{O2 + h$\nu$ &-> O + O} \tag{R1} \label{CM:R1}\\
    \ce{O2 + O &-> O3} \tag{R2} \label{CM:R2}\\
    \ce{O3 + h$\nu$ &-> O2 + O} \tag{R3} \label{CM:R3}
\end{align}
As depicted in the lower left panel of Fig. \ref{fig:UVSpecsXsecs}, \ce{O3} is also a strong UV absorber and thus further enhances this shielding phenomenon in the \PIE.  This leads to less direct photolysis of \ce{CH4}, \ce{H2O}, and HO$\mathrm{_x}$ molecules --- as well as \ce{O2} and \ce{O3} themselves --- deeper in the \PIE{} atmospheres.  In contrast, the anoxic \ARE{} lacks strong photochemical shielding via \ce{O2}, but \ce{CO2} (10\% in these atmospheres) absorbs across a similar wavelength range as \ce{O2} with a smaller cross-section.  Thus, \ce{CO2} provides weaker photochemical shielding in the \ARE, and photochemical reaction rates are higher in the \ARE's dense lower atmosphere compared to the \PIE.  These differences in the abundance and cross section of \ce{O2} and \ce{CO2} lead to the dense lower atmospheres of the \PIE{} being more compositionally stable to the evolution of M star UV flux, when compared to those of the \ARE.


\subsection{Oxygen/Ozone Buildup} \label{sec:discussion:o2o3buildup}

The abiotic buildup of \ce{O2} and \ce{O3} in the \ARE{} atmospheres around young M4 and M8 stars presented here is primarily attributed to the high abundance (10\%) of \ce{CO2}, which imparts two main effects.  First, \ce{CO2} is readily photodissociated by $\lesssim$170 nm photons (see Fig. \ref{fig:UVSpecsXsecs}) into \ce{CO} and \ce{O}, the recombination of which is spin-forbidden and slow, \citep[\ce{CO2} recombination primarily takes place via \ce{CO} + \ce{OH} in Earth's atmosphere,][]{McElroy1970_CO2photochem}, leaving free \ce{O} to bond with itself to form \ce{O2} and subsequently \ce{O3}. Second, \ce{CO2}'s absorption of FUV photons in the upper atmosphere shields other molecules, namely \ce{H2O} and \ce{CH4}, from being dissociated into reducing \ce{H} radicals that would destroy \ce{O2} and \ce{O3}. In short, a \ce{CO2} abundance of 10\% and a sufficient net UV flux and FNR allows for \ce{O2} and \ce{O3} production to be strongly favored over destruction.

The buildup of \ce{O2} and \ce{O3} in M star exoplanet atmospheres has been a subject of study throughout the literature surrounding biosignatures around these small stars \citep{Domagal-Goldman:2014_AbioO2O3, Harman:2018_AbioOxyLightning, Hu:2020}, and a comparison to these studies show our values are in agreement to varying degrees.  \citet{Domagal-Goldman:2014_AbioO2O3} shows that an anoxic Earth with 5\% \ce{CO2}, 1\Seff{} instellation from GJ 876 (M4V), and an \ce{H2} out-gassing rate equivalent to what we use in this study ($\mathrm{4 \times 10^{8}}$ molecules cm$^{-2}$ s$^{-1}$) developed an \ce{O3} column density on the order of  $16\leq\eta_{\mathrm{O3}}\leq16.5$.  Most of the \ce{O2}/\ce{O3}-buildup \ARE{} cases in this work have \ce{O3} column depths fall within or below this range ($11.7\leq\eta_{\mathrm{CH4}}\leq17.0$), with the 1 Gyr median, 650 Myr median, 650 Myr upper quartile M4 stars inducing $16.5\leq\eta_{\mathrm{CH4}}\leq17.0$.  \citet{Harman:2018_AbioOxyLightning} modeled an anoxic, 5\% \ce{CO2}, 0.6\Seff{} Earth-analogue around GJ 876, resulting in an \ce{O2} surface mixing ratio of $\mathrm{7\times10^{-13}}$, which we reproduce to within 25\% in our old (3 Gyr and 5 Gyr) M4 \ARE{} planets.  However, when comparing simulations of planets orbiting M8 stars, the 10\% \ce{CO2} TRAPPIST-1 e modeled by \citet{Hu:2020} does not agree with our results.  Their models resulted in $\mathrm{2\times10^{-7}}$ bar of \ce{O3}, about forty times what our \ARE{} builds up around our most UV emissive star ($\mathrm{5.4\times10^{-9}}$ bar of \ce{O3}).


The abiotically-generated 0.26\um{} \ce{O3} feature depicted in Figures \ref{fig:VMR_R&TSpecs_M4AgeActivity_ARE} and \ref{fig:VMR_R&TSpecs_M8AgeActivity_ARE} are strikingly similar to those produced by models of seasonal photosynthetic activity on the Proterozoic Earth \citep[e.g.  Figure 3 of ][]{Olson:2018_SeasonalOxygen}.  During that time in Earth's history, \citet{Olson:2018_SeasonalOxygen} assume an \ce{O2} surface abundance of 15 ppm, which is close to 10$\mathrm{^{-4}}$ of the present atmospheric level of \ce{O2} and consistent with geochemical proxies that constrain the \ce{O2} abundance at that time.  This small amount of \ce{O2} produces extremely weak absorption at visible wavelengths, but its photochemical proxy, \ce{O3}, produces a strong Hartley-band feature in the UV \citep{Olson:2018_SeasonalOxygen} like the one seen in this work.  Compared to the 15 ppm of \ce{O2} and the peak \ce{O3} volume mixing ratio of $\approx$10$^-9$ in the G dwarf planet of \citet{Olson:2018_SeasonalOxygen}, the \ARE{} atmospheres presented here build up to 50--100 ppm of abiotic \ce{O2} and a peak \ce{O3} volume mixing ratio of 2.9$\times$10$^8$ due to the stronger \ce{CO2}-photolyzing FUV flux from our M stars.

It will be important to characterize the planet's \ce{CO2} abundance in interpreting any \ce{O2} \& \ce{O3} signals as the higher \ce{CO2} abundances expected for planets further out in the habitable zone could provide the substrate for abiotic production of \ce{O2} and \ce{O3}.  \ce{CO2} is the major source of abiotic oxygen buildup discussed in this study, and its abundance and photolysis rate are critical to the interpretation of \ce{O2} and \ce{O3} as biosignatures.  As described in \S \ref{sec:methods:inputs:atm}, all of the model atmospheres in this study have a lower instellation (66\% S$_{eff}$) than Earth, and need significant abundances of \ce{CO2} (10\%) to retain a habitable surface temperature (284K $\pm$ 2K).  This higher abundance is consistent with carbonate-silicate cycling, a negative feedback loop between planetary surface temperature and \ce{CO2} abundance.  This process, which requires an ocean as well as active volcanism \& plate tectonics, is thought to have buffered the Earth's surface temperature over Earth's geological history \citep{Walker:1981, Kasting:1993} and is hypothesized to be able to operate to the outer edge of the HZ, where \ce{CO2} buildup past$\approx$10 bars can no longer warm the planet \citep{Kopparapu:2013}.  Consequently, although the 10\% \ce{CO2} used here is substantially larger than the true \PIE{} (280 ppm) and the value assumed for the \ARE{} (640 ppm), it is near the lower end of \ce{CO2} abundances expected for a planet within the HZ.  Thus, M star planets with different instellations and \ce{CO2} concentrations may also be vulnerable to this process \citep{Domagal-Goldman:2014_AbioO2O3, Harman:2018_AbioOxyLightning, Hu:2020, Ranjan:2022_PhotochemRunaway}, but this work illustrates how oxygen \& ozone buildup can occur, perhaps preferentially, in Earth-like atmospheres around young, UV emissive M stars that receive less than Earth's instellation.

Carbon monoxide is a potentially powerful discriminant of photochemically-produced false positive \ce{O2} and \ce{O3} \citep{Schwieterman:2016}.  The 4.6 \um{} CO feature in transit spectroscopy varies by 1.0 ppm for the M4 stars (Fig. \ref{fig:VMR_R&TSpecs_M4AgeActivity_ARE}) and 9.5 ppm for the M8 stars (Figure \ref{fig:VMR_R&TSpecs_M8AgeActivity_ARE}) over each spectral type’s CRUVE.  Given the strength of the \ce{CO} feature for the 10\% atmospheric \ce{CO2} abundance used here and the $\lesssim$ 5 ppm noise floor of JWST \citep{Lustig-Yaeger:2023_jwstLHS475b}, if the current high stellar contamination in transmission spectra \citep[eg.][]{Lim:2023_T1bStellarNoise,Radica:2025_T1cStellarNoise} can be ameliorated, \ce{CO} may be a potential false positive discriminator for abiotic \ce{O2} production, depending on a planet's atmospheric \ce{CO2} abundance and \ce{CO} production rate.

A compounding factor in the abiotic production of \ce{O3} is that the anoxic, \ce{CO2}-rich atmospheres that are most susceptible to abiotic \ce{O2} and \ce{O3} buildup may be more likely (vs. their Modern Earth counterparts) to exist around the young stars that induce this phenomenon.  On Earth, oxygenic photosynthesis is thought to have developed between 2.35 and 3.8 Gyr ago, and the Great Oxidation Event, when \ce{O2} became a major constituent of Earth's atmosphere, is estimated to have taken place between 2.3 and 2.5 Gyr ago \citep{Lyons:2014_RiseOfOxygen}.  Additionally, \citet{Catling:2020_ArcheanAtmosphere} discussed how \ce{CO2} was significantly more abundant during Earth's Archean eon, with abundances of approximately $\approx$0.1 and $\approx$0.03 bar at 4 and 2.5 billion years ago, respectively.  It is therefore plausible that an Earth-like planet in orbit of a young ($\le$1 Gyr) M star would closely resemble the anoxic and \ce{CO2}-rich \ARE{} cases presented in this study.  However, \citet{Luger:2015_OceanLoss} showed that the pre-main sequence phase of later type M stars in particular are capable of turning Earth-like planets into dessicated worlds with hundreds of bars of \ce{O2}.  The chemistry (both at the surface and in the atmospheres) of such worlds are entirely different from those modeled here, but it is plausible that hundreds of bars of oxygen produced by a late type M star could result in significant photochemical shielding (Section \ref{sec:discussion:shielding}) that restricts host star age-related effects on photochemistry to the planet's upper atmosphere.

\subsection{A Robust \ce{O2} and \ce{O3} Buildup Mechanism} \label{sec:discussion:robustO2O3}

Several studies have shown that photochemical models can produce unphysical \ce{O2} and \ce{O3} buildup if various mechanisms (e.g. redox balance) and assumptions (e.g. molecular absorption cross sections and pressure-altitude grid resolution and range) are neglected.  In this section, we show that the work we present here has considered each of these mechanisms and that the build-up of \ce{O2} and \ce{O3} driven by the host-star's UV spectrum remains robust.

The imposition of global (atmospheric + oceanic) reduction-oxidation, or ``redox'', balance is one such critical mechanism in determining the equilibrium abundances of biosignature gases like \ce{O2} and \ce{O3} \citep{Domagal-Goldman:2014_AbioO2O3, Harman:2015_AbioOxy}.  As described in Section \ref{sec:methods:inputs:atm}, each of the simulated atmospheres we present here are redox balanced to within $\lesssim$ 40 ppm and the net oxidizing flux entering the ocean from each atmosphere is well within the range of plausible reducing fluxes at the ocean mantle boundary, $F_\mathrm{floor}$.  If we had instead limited $F_\mathrm{floor}$ to the smaller values quoted in \citet{Domagal-Goldman:2014_AbioO2O3}, enforcing redox balance may have required lowering the surface deposition velocities of reducing gases such as \ce{CO}, thus accelerating the recombination of \ce{CO2} and inhibiting the build up of \ce{O2} and \ce{O3}.  Further work will be needed to determine how sensitive the \ce{O2} and \ce{O3} buildup behavior seen here is to varying assumptions about redox sources and sinks.  However, we stress that the maximum oxidizing flux entering the ocean from the atmospheres we present here is just 8\% of the estimated $F_\mathrm{floor}$ on the early Earth \citep{Holland:1984_AtmOcnEvo}.  That is, the \ARE{} ocean-mantle boundary could easily neutralize the oxidizing power from any \ce{O2} and \ce{O3} buildup if the atmosphere-ocean system was not already in redox balance.  We therefore argue that our atmospheres do not exhibit \ce{O2} and \ce{O3} buildup due to redox imbalance.

The impact of the \ce{CO2} cross-section on anoxic atmospheric chemistry was recently discussed by \citet{Broussard:2025_CO2xsec}, which showed that their ``least conservative'' \ce{CO2} cross-section resulted in significant \ce{CO2} photolysis.  This led to subsequent higher abundances of \ce{O2} across a large range of \ce{CO2} surface mixing ratios and stellar spectral types.  We use the \citet{Lincowski:2018} \ce{CO2} cross-section in this work, which is at or below the ``recommended'' cross section and well below the ``least conservative'' cross section of \citet{Broussard:2025_CO2xsec} (see their Figure 1.)  Their Figure 8 showed that the \citet{Lincowski:2018} cross section resulted in surface \ce{O2} mixing ratios that are within $\approx$0.5 dex of their ``most conservative'' cases for a 10\% \ce{CO2}, anoxic atmosphere --- akin to the \ARE{} in this work --- around GJ876 (M4V), Proxima Centauri (M6V), and TRAPPIST-1 (M8V).  We therefore argue that the \ce{O2} \& \ce{O3} buildup seen in this work is not caused by an over-estimation of the \ce{CO2} absorption cross-section.

\citet{Ranjan:2023_UpperAtmRunaway} found that photochemical models using an insufficient maximum altitude (z$_\mathrm{max}$) --- or too low an altitude grid resolution --- resulted in significant \ce{CO2} photolysis at the top-most atmospheric layer, referred to in that work as a failure to ``resolve the \ce{CO2} photolysis peak.''  This in turn drove a sharp increase in \ce{CO} and \ce{O2} abundances in their models.  Specifically, \citet{Ranjan:2023_UpperAtmRunaway} showed that, for a 0.1 p\ce{CO2} bar atmosphere like those presented in this work, a z$_\mathrm{max} =$ 54 km (minimum pressure of 0.34 $\mu$bar) was insufficient to resolve the \ce{CO2} photolysis peak in the atmosphere, while pushing z$_\mathrm{max} =$ 100 km (minimum pressure of 4 nanobar) prevented \ce{CO2} photolysis from occurring in just the top most atmospheric layer, thus avoiding the exaggerated \ce{CO} and \ce{O2} levels previously produced.  As discussed in Section \ref{sec:methods:inputs:atm}, the \ARE{} atmospheres in this work are modeled up to altitudes of 98-104 km (38-40 nanobar), depending on the incident stellar spectrum, and vertically-resolved reaction rates from our photochemical model confirmed that the model was appropriately spreading the high-altitude \ce{CO2} photolysis across the upper atmospheric layers.  We thus argue that the \ce{O2} \& \ce{O3} buildup seen in this work is not due to an incorrectly resolved \ce{CO2} photolysis curve in the upper atmosphere.

\subsection{Observational Consequences} \label{sec:discussion:observing}

We find that the majority of the changes to transmission and reflected light spectral features in our study will be difficult to detect via JWST and the next generation ELTs, respectively.  Methane's abundance increases by $\approx$36x over the CRUVE of our M4 and by a factor of 2 over the CRUVE of our M8 stars. While the variability in methane's transmission features in the M4 cases are below the approximate JWST noise floor of $\lesssim$ 5 ppm \citep{Lustig-Yaeger:2023_jwstLHS475b}, methane's transmission features in the M8 cases possess 8 and 5 ppm variability (\PIE{} and \ARE, respectively) and could potentially be detectable with JWST, though the signal would need to be disentangled from the much stronger stellar contamination \citep{Lim:2023_T1bStellarNoise, Radica:2025_T1cStellarNoise}.  Oxygen, however, will prove difficult: \citet{Lustig-Yaeger:2019_T1JWST} showed that \ce{O2}-\ce{O2} CIA bands at 1.1 and 1.3 \um{} can be detected with JWST with several bars of \ce{O2} present in the atmosphere, but the maximum \ce{O2} abundance of 0.21 bar in our atmospheres make these bands weak and the resulting features and changes to the features via UV emission are well below JWST's noise floor.  The upcoming generation of ground-based ELTs may be more suited to pursuing these features and their UV emission driven changes \cite{Snellen:2015}, but they may still require years of observation to detect for distant planets around late-type M-dwarfs \cite{Currie:2023_GroundBasedO2} like the M8 stars in our study.

While most of the UV-driven compositional changes are likely extremely challenging to detect in transmission spectra with JWST or the ELTs, the Habitable Worlds Observatory may be sensitive to the abiotically-generated UV Hartley band \ce{O3} reflectance feature, which could constitute a false positive biosignature for oxygenic photosynthesis.  The LUVOIR Mission Concept Study Final Report \citep{LUVOIR:2019} shows that an 8m LUVOIR-B-class observatory, which is potentially analogous to HWO, would be likely to detect $\approx$9 ``warm'' terrestrial exoplanets, with 3 of those expected to be ``exo-Earth'' planets, around M stars.  Our results show that this \ce{O3} spectral feature may grow stronger for anoxic, Archean-like high-\ce{CO2} planets orbiting younger ($\le$1 Gyr), more UV emissive M stars.  

The trend towards higher production of abiotic \ce{O3} around younger, more UV emissive stars is potentially applicable to earlier-type M stars and late-type K stars \citep{Arney:2019_KdwarfAdvantage} that may be prime targets for large space-based direct imaging missions.  Early-M and K stars often have substantially lower FNR \citep{Richey-Yowell:2019_KstarUV} than the late-type M stars discussed here. However, \citet{Richey-Yowell:2023_UVEvoGaia} showed that young (650 Myr) early-type M stars and K stars can exhibit $\approx$200 mW m$^{-2}$ of total habitable zone FUV+NUV flux, which is comparable to the total UV fluxes that produced the abiotic \ce{O2} and \ce{O3} buildup behavior seen in this work. Detailed stellar atmosphere and photochemical models of K stars and their planets are needed to robustly test whether these planets are susceptible to this phenomena.

\subsection{Biosignature Consequences} \label{sec:discussion:biosigs}

Without knowledge of the parent star’s UV spectrum and its capability to produce \ce{O2} and \ce{O3} in a planetary atmosphere of a given composition, discriminating between the abiotic and biogenic production of trace amounts of \ce{O3} could be challenging. For example, the abiotic \ce{O3} generated in our M4 \ARE{} simulations produced a UV feature that is comparable to that seen by \citet{Olson:2018_SeasonalOxygen} for \ce{O3} generated from small amounts of biogenic \ce{O2} on a planet developing an oxygenic photosynthetic biosphere, and planets orbiting early type M stars may be observed in the HWO target sample \citep{LUVOIR:2019, Mamajek:2024_HWOtargetList}. While the \ce{CH4} fluxes in our simulations were within Earth’s biogenic range and therefore characteristic of a methanogenic biosphere, whether the \ce{O3} signal was due to a growing photosynthetic biosphere, coincident with the methanogenesis, might be harder to interpret.

Consequently, every effort should be made to characterize the UV spectrum of the parent star and determine the presence and abundance of other atmospheric species, which could help reveal the source of the \ce{O2} and/or \ce{O3}.  One such approach is to quantify the available \ce{CO2} and look for byproducts of \ce{CO2} photolysis.  For example, several studies have suggested that abundant \ce{CO} in tandem with abundant \ce{O2} or \ce{O3} makes a strong case for extensive \ce{CO2} photolysis rather than biologic activity, although significant \ce{CO} buildup is favored in non-habitable, very low water environments \citep{Gao:2015_CO2Stability, Schwieterman:2016}.  However, seasonal variability enhanced by a non-homogeneous distribution of land may result in changing strength of the UV \ce{O3} feature with orbital position \citep{Olson:2018_SeasonalOxygen} which could provide an additional discriminant.  For a circular orbit, similar variability would not be expected for a purely atmospheric, abiotic signal, although an elliptical orbit may produce similar behavior to the biogenic case and would need to be ruled out.  A strong reduction in obliquity via tidal locking \citep{Heller:2011_TidalObliq} could also weaken the seasonality of the biological \ce{O2} and \ce{O3} signal, although planet-planet interactions between closely-packed M star planets could maintain a non-zero obliquity \citep{Ribas:2016,Meadows:2018_PCb, Bolmont:2015_Kepler62TidalEvo} that may still permit some seasonality.

Like previous work by \citet{Segura:2005_MdwarfBiosig}, we find that methane concentrations are higher in Earth-like atmospheres orbiting M stars, but also that methane's strong dependence on the total incident UV flux (\S \ref{sec:results:FNRvsNetUV:ch4}) and falloff of M star UV flux with age results in higher methane abundances in these atmospheres than previously modeled. \citet{Segura:2005_MdwarfBiosig} found that a modern Earth-like planet is capable of building up significant amounts of surface methane when orbiting AD Leo (400--500 ppm) compared to an Earth-Sun analogue (1-2 ppm).  The \PIE{} cases we present here are indeed rich in methane compared to a preliminary \PIE{} + Sun photochemical model (0.6 ppm), but surface methane mixing ratios reach well above 1000 ppm around the older, less UV-emissive M stars included in this study.  \citet{Segura:2005_MdwarfBiosig} describes how their increased surface abundance of \ce{CH4} is primarily due to a significant reduction in the methane-destroying \ce{OH} radical, which is indirectly produced in Earth's atmosphere via photochemistry driven by 200--300 nm radiation.  The AD Leo spectrum used in that work insolated their Earth-like planet with $\approx$100 mW m$^{-2}$ of ``UV-C'' (200-280 nm) radiation, while the lowest UV activity star included here (lower quartile 5 Gyr M4) insolates the \PIE{} with only 11.5 mW m$^{-2}$, about 10\% of the AD Leo insolation. This leads to the \PIE{} + lower quartile 5 Gyr M4 case building up $\approx$2700 ppm of methane at the surface. Thus, our work suggests that the increase of methane in Earth-like atmospheres around M stars may be stronger than previously thought, and is a strong function of age and UV emission from the host M star.\newline

\subsection{Characterizing M Star UV Flux} \label{sec:discussion:uvchar}


Our study shows that the abundances of several key biosignature gases are dependent on photochemistry driven by both net UV flux and FNR  of the host M star's UV spectrum (\S \ref{sec:results:FNRvsNetUV}), and multi-modal characterization of M star UV spectra will be needed to interpret the presence (or lack thereof) of a range of biosignature gasses.

As discussed above, the correlations between M star UV variability and biosignature gas abundance may change with different atmospheric compositions and boundary conditions, but several other studies have shown that specific characteristics of the host star's UV spectrum impact specific planetary atmospheric characteristics.  \citet{Teal:2022_UncertaintyUV} showed that haze formation and the spectral feature strength due to hazes in \ARE-like atmospheres were strongly dependent on the host-star's UV continuum.  We do not include haze production in our \ARE{} cases, so further work will be necessary to understand how the variance of UV flux from M stars across stellar age impacts photochemical haze production.  Stellar flares are also a strong source of UV flux from M stars \citep{Mirzoian:1980_FlareStars, Loyd:2018}, and the effects of both singular flare events \citep{Segura:2010_MdwarfFlarePhotochem} and repeated flaring \citep{Tilley:2019} on terrestrial worlds around M stars have shown that while radiation-only events cause negligible change in atmospheric ozone abundance, radiation + proton events can cause a $\ge$90\% reduction in the \ce{O3} column depth of an Earth-like atmosphere.  We note that the results shown in this study are products of only quiescent UV emission, and the impact of differences in flare rates and energies across stellar age on terrestrial atmospheres will need further study.


Lastly, we discuss the implications of using the \citet{Peacock:2019_T1} TRAPPIST-1 spectra in this work in lieu of the more recent work of \citet{Wilson:2021_T1MegaMUSCLES}.  For a \PIE-like TRAPPIST-1 e, \citet[][see their Fig. 1 and Section 2.1]{Cooke:2023_T1UVO3} showed that \ce{O3} abundances increased by a factor of 26 (1.4 dex) when the planet was irradiated with the \citet{Peacock:2019_T1} 1A spectrum compared to the UV spectrum of \citet{Wilson:2021_T1MegaMUSCLES} due to the \citet{Peacock:2019_T1} spectra over-predicting emission line fluxes that the noiseless \citet{Wilson:2021_T1MegaMUSCLES} spectrum was modeled to match.  Far-UV emission lines (e.g Si III, Fe II, Si II, and Al I) are between $-$1 and 3.7 dex lower/higher in the \citet{Peacock:2019_T1} spectrum used in this work compared to the \citet{Wilson:2021_T1MegaMUSCLES} spectrum, and as such likely play an important role in the photolysis of \ce{CO2} seen in the \ARE{} atmospheres presented here.  However, while emission lines make up the majority of integrated flux in the FUV, the \citet{Wilson:2021_T1MegaMUSCLES} spectrum only matched the \ce{Mg} II (2796 \Ang) emission line in the NUV, thus underestimates the star's NUV spectrum compared to WCF3 NUV photometry (Wilson, private communication). This results in a GALEX FNR of 0.39 for the \citet{Wilson:2021_T1MegaMUSCLES} spectrum.  In contrast, the \citet{Peacock:2019_T1} spectra used in this work are based on a statistical, empirically-guided stellar model that includes a physical treatment of the UV continuum and are consistent with the GALEX FNR of 0.26 of a field-age M8.5 star.  While the M8 spectra used in this work might not be entirely representative of TRAPPIST-1, variations between individual late M stars at field ages vary across an order of magnitude in NUV flux and over two orders of magnitude in FUV flux \citep{Schneider:2018_HAZMAT3}.  We stress that the M8 spectra we have used here best represent the population average of late M stars and not any single individual star such as TRAPPIST-1.  We showed in this work that the abundance of \ce{O3} in the \ARE{} is dependent on both the net UV flux incident on the planet and the FNR of the UV spectrum.  Should this study be repeated using the current \citet{Wilson:2021_T1MegaMUSCLES} spectrum, we expect that the lower net UV flux --- especially from FUV emission lines --- would push the atmospheres towards less \ce{O2} and \ce{O3}, but this effect would be balanced by the \citet{Wilson:2021_T1MegaMUSCLES} spectrum's higher FNR pushing the atmosphere towards more \ce{O2} and \ce{O3}.  Further work will be needed to disentangle the impact of these opposing properties of the UV spectrum on the abundances of \ce{O2} and \ce{O3} in Earth-like atmospheres.

\section{Conclusion} \label{sec:conclusion}

Using coupled climate/photochemical models we have explored the impact of age-dependent changes in  the host stellar UV spectrum and flux on the composition of the atmospheres of Earth-like planets orbiting M4 and M8 dwarfs.  We find that the heightened total UV flux and higher FUV/NUV ratios (FNR) of young (650 Myr and 1 Gyr old) M stars drive a more than 5 dex increase in the abundance of \ce{O3} and a 1.6 dex decrease in the abundance of \ce{CH4} compared to older (5 Gyr) M stars.  Subsequent changes in the spectral features of some of these gases are potentially detectable with current transit-transmission and next-generation direct-imaging spectroscopic missions.

Our study shows that Earth-like planets orbiting M stars may be more susceptible to abiotic oxygen and ozone buildup if the host star is younger than $\approx$1 Gyr, especially for \ce{O2}-poor, \ce{CO2}-rich atmospheres that may be characteristic of young planets without oxygenic biospheres.  Our \PIE{} atmospheres have enough biological \ce{O2} (21\%) that significant photochemical shielding occurs and the impact of each star's UV emissivity range (\S\ref{sec:methods:inputs:activityrange}) on the abundances of atmospheric \ce{O3} and \ce{CH4} is reduced.  Our \ARE{} atmospheres are more susceptible to stellar UV emissivity as they are relatively anoxic, but still possess 10\% \ce{CO2} that provides weaker photochemical shielding.  Thus, heightened UV emission when the host M star is young ($\lesssim$1 Gyr) drives extensive \ce{CO2} photolysis in the \ARE{} and the accumulation of up to 90 ppm of \ce{O2} at the planet surface.  While this amount of \ce{O2} produces only very weak spectral features in the visible, \ce{O3} --- a photochemical byproduct --- reaches concentrations sufficient to produce a pronounced Hartley band (0.26\um) feature in UV reflectance spectroscopy.  These changes in the strength of the Hartley band feature due to the age and UV-emissivity of the host M star may be detectable with the Habitable Worlds Observatory, making stellar-age-dependent UV emission a critical piece of context in the detection and interpretation of \ce{O3} as a biosignature, especially for early-Earth-like environments.  Similarly, characterizing the ages of potential target systems will be crucial in designing precursor studies for the Habitable Worlds Observatory.

We find that methane's photochemical lifetime increases as stellar age increases from 650 Myr to 5 Gyr so that, for the same \ce{CH4} surface flux, the abundance of \ce{CH4} increases by up to $\approx$36x (M4 + \ARE cases). This behavior is inversely related to the abundances of \ce{O2} and \ce{O3} in our atmospheres.  The increase in methane's photochemical lifetime is mainly due to a) weakening of direct \ce{CH4} photolysis, and b) fewer reactions with the destructive radicals \ce{OH} (via water photolysis) and \OsD{} (via photolysis of \ce{O2}, \ce{O3}, and \ce{CO2}) as the host star ages.  The concentration of \ce{CH4} increases by, at maximum, $\approx$ 1.5 dex and it is possible that the maximum strengthening of its spectral features ($\approx$8 ppm) could be resolved with JWST \citep[5 ppm noise floor,][]{Lustig-Yaeger:2023_jwstLHS475b}, though this signal would need to be disentangled from much stronger stellar contamination \citep{Lim:2023_T1bStellarNoise, Radica:2025_T1cStellarNoise}.  Therefore, our results suggest it may be possible that stellar-age-dependent UV emission from M stars could impact future work in constraining atmospheric methane abundances of planets similar to those presented here via JWST.

Finally, we show that the abundances of two key biosignature gasses, \ce{O3} and \ce{CH4}, in the \ARE{} are strongly correlated with the net UV flux and/or FNR of their host M star's UV spectrum.  Ozone abundances are strongly tied to both the net UV flux and the UV spectral shape of the host star, while methane abundances are primarily correlated with net UV flux with little discernible dependence on the shape of the UV spectrum.  Thus, characterizing these two properties of a M star's UV spectrum will be critical to interpreting the abundance of these two biosignature gases in Earth-like atmospheres around these small stars.

\section{Acknowledgments} \label{sec:acknowledgements}

The authors thank an anonymous reviewer for their pertinent comments that strengthened this work and its conclusions. We thank Shawn Domagal-Goldman and Chester E. Harman for useful discussion on the sources \& sinks of molecular oxygen and ozone, Jacob Lustig-Yaeger for advice on estimating detectabiltiy, Mike L. Wong for his expertise in disentangling photochemical reaction networks, and Miles H. Currie for discussions on the detectability of M-dwarf planets with ELTs, all of which proved invaluable to this study.  Lastly, we thank XKCD and their RGB color survey for lending a splash of color to our figures.  This work was performed by the Virtual Planetary Laboratory Team, a member of the NASA Nexus for Exoplanet System Science, funded via NASA Astrobiology Program Grant Nos. 80NSSC18K0829 and 80NSSC23K1398.  E.S. and R.O.P.L. also appreciate NASA support through a grant from the Space Telescope Science Institute (HST-GO-14784.001-A), which is operated by the Association of Universities for Research in Astronomy, Inc., under NASA contract NAS 526555.  S.P. and E.S. also acknowledge support from the CHAMPs (Consortium on Habitability and Atmospheres of M-dwarf Planets) team, supported by the National Aeronautics and Space Administration (NASA) under grant nos. 80NSSC21K0905 and 80NSSC23K1399 issued through the Interdisciplinary Consortia for Astrobiology Research (ICAR) program.  S.P. also acknowledges support from NASA under award number 80GSFC24M0006.  This work made use of the advanced computational, storage, and networking infrastructure provided by the Hyak supercomputer system at the University of Washington.

\bibliographystyle{aasjournal} \bibliography{bib}



\end{document}